\def\etal{{et al. }}

\def\keV{{\rm keV}}

\def\Omat{\Omega_{\rm M}}
\def\Olam{\Omega_{\Lambda}}

\documentclass[preprint2]{aastex}
\usepackage{epsfig}
\begin{document}
%\lefthead{VOIT ET AL.}
%\righthead{INTRACLUSTER ENTROPY}
\shorttitle{INTRACLUSTER ENTROPY}
\shortauthors{VOIT ET AL.}
\slugcomment{\apj , submitted 18 Dec 2002}
\title{On the Origin of Intracluster Entropy}
\author{G. Mark Voit\altaffilmark{1},
        Michael L. Balogh\altaffilmark{2},
        Richard G. Bower\altaffilmark{2}, \\
        Cedric G. Lacey\altaffilmark{2,3}, \&
        Greg L. Bryan\altaffilmark{4}
         } 
\altaffiltext{1}{Space Telescope Science Institute,
                 3700 San Martin Drive, 
                 Baltimore, MD 21218, 
                 voit@stsci.edu}
\altaffiltext{2}{Department of Physics,
                 University of Durham, 
                 South Road,
                 Durham DH1 3LE, UK, 
                 M.L.Balogh@durham.ac.uk, 
                 R.G.Bower@durham.ac.uk}
\altaffiltext{3}{Observatoire de Lyon,
                 9 Avenue Charles Andre, 
                 69230 Saint Genis Laval,
                 France, 
                 Cedric.Lacey@obs.univ-lyon1.fr}
\altaffiltext{4}{Physics Department,
                 University of Oxford, 
                 Keble Road, 
                 Oxford OX1 3RH, UK, 
                 gbryan@astro.ox.ac.uk}

\begin{abstract}
The entropy distribution of the intracluster medium 
and the shape of its confining potential well completely
determine the X-ray properties of a relaxed cluster of
galaxies, motivating us to explore the origin of 
intracluster entropy and to describe how it develops 
in terms of some simple models.  
We present an analytical model for smooth accretion, including
both preheating and radiative cooling, that links a cluster's
entropy distribution to its mass accretion history and shows 
that smooth accretion overproduces the entropy observed in
massive clusters by a factor $\sim 2$-3,
depending on the mass accretion rate.
Any inhomogeneity in the accreting gas reduces entropy
production at the accretion shock;  thus, smoothing of
the gas accreting onto a cluster raises its entropy level.
Because smooth accretion produces more entropy than 
hierarchical accretion, we suggest that some of the 
observed differences between clusters and groups may 
arise because preheating smooths the smaller-scale lumps 
of gas accreting onto groups more effectively than it
smooths the larger-scale lumps accreting onto clusters.  
This effect may explain why entropy levels at the outskirts 
of groups are $\sim 2$-3 times larger than expected from self-similar
scaling arguments.  The details of how the density
distribution of accreting gas affects the entropy
distribution of a cluster are complex, and we suggest
how to explore the relevant physics with numerical simulations.
\end{abstract}


\setcounter{footnote}{0}

\section{Introduction}

Not so very long ago, people who studied clusters of galaxies
were often asked how the X-ray emitting gas gets so hot.  The
answer is simple, of course.  If radiative cooling is negligible,
then gravitationally driven processes will heat diffuse gas to the 
virial temperature of the potential well that confines it.
A tougher question would have been to ask why the intracluster
medium has the density that it does.  In order to answer
that question, one needs to know what produces the entropy
of the X-ray emitting gas.

Entropy is of fundamental importance because a cluster's 
intergalactic gas will convect until its isentropic surfaces 
coincide with the equipotential surfaces of the dark-matter 
potential.  Thus, the entropy distribution of a cluster's 
gas and the shape of the dark-matter potential well in which 
that gas sits completely determine the large-scale X-ray 
properties of a relaxed cluster of galaxies (see Voit \etal 
2002 and references therein).  The gas density profile $\rho(r)$ 
and temperature profile $T(r)$ in this state of convective and 
hydrostatic equilibrium are just manifestations of the 
underlying entropy distribution.  
If we wish to link these X-ray observables to the process 
of cluster formation, we therefore need to understand how 
the growth of cosmic structure generates intracluster entropy 
and how processes like radiative cooling and non-gravitational 
heating by supernovae and active galactic nuclei modify 
that entropy distribution.

One way to approach the problem of gravitationally driven
entropy generation is through spherically symmetric
numerical models of smooth accretion, in which gas passes through 
an accretion shock as it enters the cluster (e.g., Knight \& Ponman
1997; Tozzi \& Norman 2001).
If the incoming gas is cold, then the accretion shock is the
sole source of intracluster entropy.  If instead the incoming gas has
been heated before passing through the accretion shock,
then the Mach number of the shock is smaller and the intracluster
entropy reflects both the amount of preheating and the production
of entropy at the accretion shock. 

In reality, however, the accreting gas is lumpy, not smooth.  
Incoming gas associated with accreting subhalos enters the 
cluster with a wide range of densities.  There is no well-defined 
accretion shock but rather a complex network of shocks as 
different lumps of infalling gas mix with the intracluster medium 
of the main halo.  Yet, despite this complexity, numerical simulations 
of hierarchical structure formation show that the resulting 
entropy profile is similar to that found in the smooth 
accretion models (Borgani \etal 2001, 2002).

This paper outlines some simple analytical models designed to
clarify the processes that determine the entropy of intracluster
gas.  As has become customary in this field, we will refer
to 
\begin{equation}
 K \; = \; \frac {P} {\rho^{5/3}} 
   \; = \; \frac {1} {\mu m_p} \left( \frac {n_e} {\rho} \right)^{2/3}
       \frac {T} {n_e^{2/3}}
\end{equation}
as the ``entropy'' of the gas, while recognizing that the 
formal thermodynamic entropy per particle for a gas of 
non-interacting monatomic particles is 
\begin{equation}
 s = \ln K^{3/2} + s_0 \; \; ,
\end{equation}
where $s_0$ depends only on fundamental constants and the 
mixture of particle masses.
Because we express temperatures in energy units throughout the
paper, Boltzmann's constant is absorbed into $T$ and $s$
becomes a dimensionless quantity.   
The object of our investigation is to understand the origin 
of the intracluster entropy distribution $K(M_g)$, 
defined so that the inverse function $M_g(K)$ is 
the mass of gas with entropy $<K$.  

Section~\ref{sec-smoothcold} computes the entropy distribution arising 
from smooth spherical accretion of cold gas.  It presents a simple 
analytical formulation for smooth accretion relating the entropy 
distribution of a cluster directly to its mass accretion rate.  The shape 
of the resulting entropy distribution is similar to that of simulated 
and observed clusters but its normalization is too large by a 
factor $\sim$2-3, indicating that inhomogeneities in the accreting 
gas must be taken into account.  Because of this difference between
smooth accretion and inhomogeneous accretion, the normalization of
the intracluster entropy profile reflects the lumpiness of the
gas that accreted onto the cluster.

Section~\ref{sec-smoothpre} shows how preheating and radiative cooling 
change the entropy distribution produced by smooth accretion.  We 
demonstrate that modest amounts of preheating raise the entropy 
distribution expected from cold accretion by an additive term 
proportional to the initial entropy of the incoming gas.  However, 
large amounts of preheating, comparable to the characteristic entropy
of the halo, suppress entropy production because they expand
the intracluster medium and reduce the shock velocity at the
accretion front.  Including simple corrections for preheating
and radiative cooling yields analytical smooth-accretion
models whose entropy distributions agree well with the
numerical models of Tozzi \& Norman (2001).  However, the
effects of preheating on hierarchical accretion are qualitatively
different because preheating smooths the gas accreting onto a 
cluster, potentially boosting its postshock entropy much more
than the simple additive correction applied to smooth accretion.

Section~\ref{sec-evidence} presents evidence suggesting that accretion of
baryons onto groups was smoother than accretion onto clusters.
We show that groups must have significant entropy gradients.
Otherwise, the observed values of core entropy cannot be
reconciled with the observed X-ray luminosity-temperature ($L$-$T$)
relation.  Polytropic models consistent with both the core 
entropy and the $L$-$T$ relation imply that entropy levels
in the outer parts of groups are $\sim$2-3 times higher 
than expected from simulations without cooling or preheating 
and from self-similar scaling of clusters.  These findings
are consistent with existing observations of groups and
suggest a transition from lumpy accretion to smooth accretion
below a mass scale $\sim 10^{14} \, h^{-1} \, M_\odot$.

Section~\ref{sec-puzzle} explores how the lumpiness of accreting
gas determines the intracluster entropy distribution.
We present a naive calculation applying simple accretion
shocks to discrete accreting subhalos and show that
this simple picture fails to produce enough entropy.  
We then consider what adjustments to the preshock density 
and preshock velocity would be needed to produce the 
proper amount of entropy through simple accretion shocks.
However, the situation could well be more complicated
than this because dense accreting lumps do not necessarily 
thermalize all their incoming kinetic energy within the
accreting gas.  We therefore generalize the idea of 
an accretion shock and investigate entropy production 
by dissipation of turbulence and relatively weak shocks 
created as accreting subhalos circulate within the 
intracluster medium.  Somehow the process of hierarchical 
accretion in the absence of preheating and radiative 
cooling produces self-similar entropy profiles in 
groups and clusters, and we assess the amount of heat 
input needed to preserve this self-similarity.  If this 
heating mode is significant, then it can partially offset
radiative cooling in the cores of clusters, and we
suggest how to test for this effect using numerical
simulations.

Section~\ref{sec-summary} summarizes our findings.

\section{Smooth Accretion of Cold Gas}
\label{sec-smoothcold}

Let us first consider the case of cold accreting gas,
in which the pressure of the incoming gas is
negligible.  We will assume a spherically symmetric
geometry, so that mass accretes in a series of
concentric shells, each with baryon fraction $f_b$,
that initially comove with the Hubble flow.
In this simple model, a shell that initially encloses 
total mass $M$ reaches zero velocity at the turnaround 
radius $r_{\rm ta}$ and falls back through an accretion
shock at radius $r_{\rm ac}$.  We will find that the entropy
distribution in this case is determined primarily
by the rate at which matter accretes onto the cluster
and yields an entropy distribution between $K \propto
M_g$ and $K \propto M_g^{4/3}$.

\subsection{Postshock Entropy}

Because the cold accreting gas is effectively pressureless, 
the equations that determine the postshock entropy are
\begin{eqnarray}
  \dot{M}_g & = & 4 \pi r_{\rm ac}^2 \rho_1 v_{\rm ac}  \\
  v_{\rm ac}^2 & = & \frac {2GM \xi} {r_{\rm ac}}  \\
  \xi & = & 1 - \frac {r_{\rm ac}} {r_{\rm ta}}  \label{eq-xi} \\
  T_2 & = & \frac {1} {3} \mu m_p v_{\rm ac}^2  \label{eq-tps} \\
  \rho_2 & = & 4 \rho_1 \; \; \label{eq-rhops} ,
\end{eqnarray}
where $\rho_1$ is the preshock gas density and 
$T_2$ and $\rho_2$ are postshock quantities.
Equations (\ref{eq-tps}) and (\ref{eq-rhops}) are restatements
of the jump conditions for strong shocks, assuming that the
postshock velocity is negligible in the cluster rest frame
(e.g., Landau \& Lifshitz 1959; Cavaliere, Menci, \& Tozzi 1997), 
and equation (\ref{eq-xi}) is strictly true only for 
cosmologies with $\Lambda = 0$.
The postshock entropy produced by smooth accretion of
cold gas at time $t$ is then
\begin{eqnarray}
 K_{\rm sm} & = & \frac {v_{\rm ac}^2} {3 (4 \rho_1)^{2/3}} \\ 
      ~     & = & \frac {1} {3} 
 		\left( \frac {4 \pi G^2 \xi^2} {f_b} \right)^{2/3}
                \left[ \frac {d \ln M} {d \ln t} \right]^{-2/3} 
                (Mt)^{2/3}
              \nonumber \; \; ,
  \label{eq-kcold}
\end{eqnarray}
where $f_b = 0.02h^{-2}\Omat^{-1}$ is the universal baryon fraction.
Thus, the entropy profile arising from smooth accretion of cold gas 
depends entirely on the ratio $r_{\rm ac}/r_{\rm ta}$ and 
the accretion history $M(t)$.

\subsection{Shock Radius}
\label{sec-accrad}

The ratio of the shock radius to the turnaround radius 
should remain nearly constant with time in the case of 
cold accretion.  One standard approach to estimating the
virial radius of a cluster is to assume that it
is precisely equal to half the turnaround radius
of the shell that is currently accreting.  Setting
the shock radius equal to the virial radius defined
in this way implies $\xi = 0.5$ by definition.

More generally, one can assume that the shock
occurs at the radius $r_{\Delta}$ within which
the mean density is $\Delta$ times the critical
density $\rho_{\rm cr}$.  In that case, $r_{\rm ac} = 
(2GM/H^2 \Delta )^{1/3}$, where $H = H_0 
[\Omat (1+z)^3 + (1-\Omat)]^{1/2}$ in a flat
cosmology.  The corresponding turnaround radius 
can be found from the equation of motion
\begin{equation}
   \ddot{r} = - \frac {GM} {r^2} + \frac {\Lambda r} {3} \; \; .
\end{equation}
In the limit of a vanishingly small cosmological constant 
$\Lambda$, a bound shell obeys the familiar parametric 
solution $r = r_{\rm ta} [(1 - \cos \theta_M) /2 ]$, 
$t = t_{\rm vir} [(\theta_M - \sin \theta_M)/ 2 \pi]$, with
$r_{\rm ta} = [(2GMt_{\rm vir}^2) / \pi^2]^{1/3}$ for a shell 
that collapses to the origin at time $t_{\rm vir}$.  The solution
for $\Omega_\Lambda = \Lambda / 3 H_0^2 = 0.7$ is not much
different because the $\Lambda$ term is always small.
The quantity $\Lambda r / 3$ is never greater than
$0.14 GM / r^2$ during the trajectory of any shell that has
accreted by the present time, which is why we neglected any
$\Lambda$-dependence in equation~(\ref{eq-xi}).
The time to fall from $r_{\rm ac} \lesssim 0.5 r_{\rm ta}$ 
to the origin is also negligible, so we obtain
$r_{\rm ac} / r_{\rm ta} \approx (\pi^2 / H^2 t^2 \Delta)^{1/3}$.
In this paper, we will generally set $r_\Delta = r_{\rm ta}/2$,
unless stated otherwise.
%Setting $r_{\rm ac} = r_{200}$ then gives $\xi \approx 0.52$
%when the matter density of the universe is nearly
%critical and $\xi \approx 0.63$ at $z = 0$ for
%$\Omat = 0.3$.

The self-similar solution of Bertschinger (1985) for 
cold accretion with $\Omega_{\rm M} = 1$ and $f_b \ll 1$
provides some support for these assumptions.
In that model, the radius of the accretion
shock remains fixed at 0.347 times the radius of
the shell that is currently turning around. 
Because $M \propto t^{2/3}$ in the Bertschinger solution
and the shell turning around at time $t$
accretes at time $2t$, this model implies
$\xi = 1 - 0.347 \cdot 2^{8/9} = 0.36$.

In a self-consistent model of a shock-bounded intracluster
medium, the ram pressure of infalling gas at the accretion 
shock must balance the thermal pressure at the outer boundary
of the hydrostatic region (i.e., $P + \rho v^2$ must be continuous
across the accretion shock.)  Thus, the value of $\xi$ depends
on both the accretion rate of the cluster and its internal
structure.  Appendix~A develops self-consistent solutions
for the equilibrium shock radius in the case of a polytropic
equation of state and a time-varying accretion history.  As 
long as the accreting gas is cold, we find that $\xi \approx
0.35 - 0.6$ for the accretion histories considered in this
paper.  However, preheating of the accreting gas can drive
$\xi$ much lower, as we will discuss in \S~\ref{sec-preheat}.

\subsection{Accretion History}
\label{sec-acchist}

Because of the constancy of $\xi$, the entropy distribution
for cold, smooth accretion directly reflects the accretion history
$M(t)$.  A rough estimate of $M(t)$ can be obtained
from extended Press-Schechter theory (Bond \etal 1991;
Bower 1991; Lacey \& Cole 1993).  If we
restrict our attention to virialized halos that are 
much more massive than the characteristic mass, then
$M \propto \omega^{-3/(n+3)}$, where $n$ is the power-law 
slope of the perturbation spectrum and $\omega \equiv \delta_c(t) 
D(t_0) / D(t)$ is a function of the critical threshold for 
virialization $\delta_c(t)$ and the growth function $D(t)$ 
(e.g., Lacey \& Cole 1993; Voit \& Donahue 1998).  In a 
flat cosmology with $\Omega_{\rm M} = 0.3$ at $t_0$, the 
approximation $D(t) \propto t^{0.63}$ is accurate to within 6\% 
from $0.01 t_0$ to $t_0$.  For the relevant range of power-spectrum 
indices ($-2 \lesssim n \lesssim -1$), we 
therefore find $M \propto t^\zeta$, with
$0.9 \lesssim \zeta \lesssim 1.9$, yielding a power-law
entropy distribution between $K_{\rm sm} \propto M_g$
and $K_{\rm sm} \propto M_g^{1.4}$.  These simple
relations agree well with those found by both spherically
symmetric hydrodynamical calculations (Tozzi \& Norman 2001)
and three-dimensional simulations (Borgani \etal 2001, 2002).
However, to do a proper comparision, we need a more
accurate expression for $M(t)$.

% ----------------------------------------
\begin{figure}[t]
\includegraphics[width=3in]{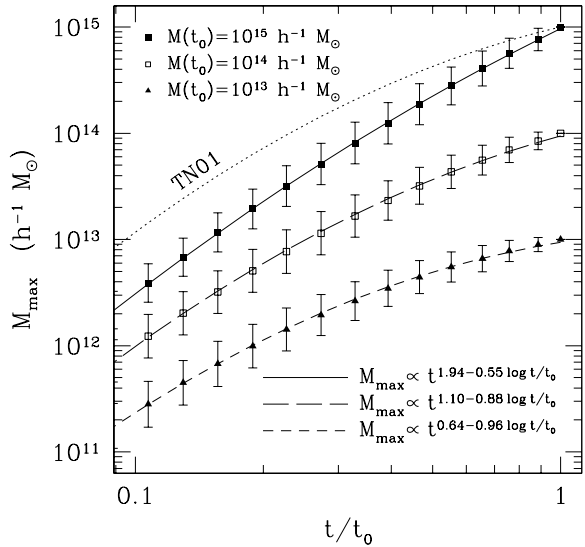}
\caption{ \footnotesize
Accretion histories for halos of total mass $M(t_0)$ at 
the present time $t_0$.  Filled squares give the logarithmic mean 
of the maximum progenitor mass $M_{\rm max}$ in the merger 
history of a halo of mass $M(t_0) = 10^{15} h^{-1} M_\odot$, 
computed from 1000 realizations of the Lacey \& Cole (1993) 
algorithm.  Open squares and filled triangles show the
corresponding quantity for $M(t_0)= 10^{14} h^{-1} M_\odot$
and $10^{13} h^{-1} M_\odot$.  Error bars indicate one 
standard deviation.  The solid line shows the best fitting
parabola in $\log M(t)$-$\log t$ space for $M(t_0) = 10^{15} 
h^{-1} M_\odot$.  The long and short-dashed lines show best fits for
$M(t_0) = 10^{14} h^{-1} M_\odot$ and $10^{13} h^{-1} M_\odot$,
respectively.  The dotted line indicates the accretion
history used by Tozzi \& Norman (2001) for their 
$10^{15} \, h^{-1} \, M_\odot$ cluster in a $\Lambda$CDM 
cosmology.
\label{mtfits}}
\end{figure}
% ----------------------------------------

\begin{figure}[t]
\includegraphics[width=3in]{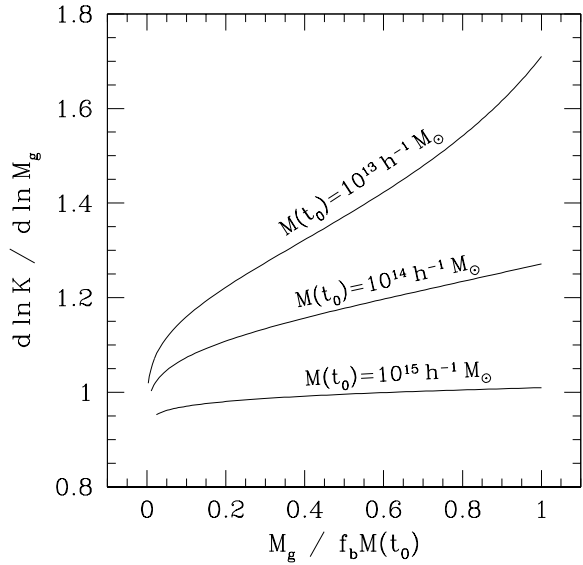}
\caption{ \footnotesize
Logarithmic slope $d \ln K / d \ln M_g$ of the entropy
profile $K_{\rm sm}(M_g)$.  The profile for a $10^{15} \, h^{-1}
\, M_\odot$ halo given by equation~(\ref{eq-kcold}) is
nearly linear, while the profiles for lower-mass halos
steepen near the outskirts of the cluster.  This steepening
arises because of the diminishing accretion rate at late 
times.  
\label{kmeffs}}
\end{figure}

Figure~\ref{mtfits} illustrates some useful expressions
for $M(t)$ derived from the merger-tree algorithm of 
Lacey \& Cole (1993) for a $\Lambda$CDM cosmology with
$\Omega_M = 0.3$, $\Omega_\Lambda = 0.7$, and $\sigma_8
= 0.9$.  We computed 1000 realizations
of merger trees leading to a present day halo mass
$M(t_0)$, assumed that $M(t)$ was equal to
the maximum progenitor halo mass $M_{\rm max}$ at time $t$,
and determined the best-fitting parabola in $\log M$--$\log t$
space.  The coefficients of those best fits are given
in Figure~\ref{mtfits}.  As expected, the effective value
of $\zeta$ is in the range $1 \lesssim \zeta \lesssim 2$,
except for the lower-mass halos ($10^{13} h^{-1} M_\odot$) 
at late times.  Figure~\ref{kmeffs} shows the logarithmic slope 
$d \ln K / d \ln M_g$ implied by these fits.  
Note that this slope generally remains between 1.0 and 1.3
but rises above that in the outer parts of lower-mass
halos because of the diminishing accretion rate as
$t$ approaches $t_0$.

The near linearity of the relation between $K$ and $M_g$
is what sets the effective polytropic index of intracluster
gas outside the cluster core.  Because the underlying 
potential is nearly isothermal, the gas density scales 
approximately as $\rho \propto r^{-2}$, implying
$M_g \stackrel {\propto} {\sim} \rho^{-1/2}$.
Thus, the relation $K \propto M_g$ leads to
$P \propto \rho^{\gamma_{\rm eff}}$ with 
$\gamma_{\rm eff} \approx 1.2$ and $K \propto M_g^{1.2}$ 
leads to $\gamma_{\rm eff} \approx 1.1$.  Observations
indicate that $\gamma_{\rm eff} \approx 1.1-1.2$ in
hot clusters (Markevitch \etal 1998, 1999; 
Ettori \& Fabian 1999; DeGrandi \& Molendi 2002), 
supporting the idea that $K \stackrel {\propto} {\sim} M_g$ 
when gravitationally-driven processes dominate the production 
of intracluster entropy.
 
% ----------------------------------------
% ----------------------------------------

\subsection{Entropy Profile}

We can now compare the entropy profile produced by
smooth accretion of cold gas to entropy profiles computed
in other ways.  To simplify those comparisions, we 
will recast the entropy profiles in dimensionless form.
Because the mean matter density within $r_\Delta$
is $\Delta \rho_{\rm cr}$, the characteristic
temperature associated with overdensity $\Delta$
is $T_\Delta = G M_\Delta \mu m_p / 2 r_\Delta$, 
where $M_\Delta = 4 \pi r_\Delta^3 \Delta \rho_{\rm cr} / 3$.
These definitions lead to a characteristic entropy 
in the baryons of $K_\Delta = T_\Delta / [\mu m_p 
(\Delta f_b \rho_{\rm cr})^{2/3}]$.  In this section, 
we divide all profiles by the cosmology-independent 
entropy scale $K_{200}$ obtained by setting $\Delta = 
200$.  The dimensionless entropy is then $\hat{K} 
\equiv K/K_{200}$, with $K_{200} = (3.3 \times 10^{34} 
\, {\rm erg \, cm^2 \, g^{-5/3}}) (M_{200}/{10^{15} \, h^{-1} \, 
M_\odot})^{2/3}$ for $\Omat = 0.3$.  In section~\ref{sec-puzzle}
we will find it more useful to work with the characteristic
entropy $K_\phi$ and characteristic temperature $T_\phi$
obtained by setting $r_\Delta = r_{\rm ta} / 2$, but comparisons
involving observations and differing cosmologies are
simpler with the more definite quantities associated with
$\Delta = 200$.

Because we are idealizing the accretion history as a smooth
increase in $M(t)$, there is a one-to-one relationship between
the entropy $K$ of a gas shell in the final configuration,
the gas mass $M_g$ enclosed within that shell, and the time
$t$ that the shell accreted.  Defining a dimensionless
accretion history $\eta(t) = M(t)/M_{200}$ therefore
allows us to write the dimensionless entropy profile as
\begin{eqnarray}
 \hat{K}_{\rm sm}(\eta) & = & 2 \left( \frac {100} {3} \right)^{1/3}
                      (H_0 t_0)^{2/3}  \, \xi^{4/3} \; \; \; \nonumber \\
 ~ & ~ & \; \; \, \times     
                 \left[ \frac {d \ln \eta} {d \ln t} \right]^{-2/3}
                 \left[ \frac {\eta t(\eta)} {t_0} \right]^{2/3} \, ,
\label{eq-khatcold}
\end{eqnarray}
with $\eta = f_g \equiv M_g/f_b M_{200}$, so that $f_g$ is the
fraction of a cluster's baryons with entropy $< \hat{K}$. 

% ----------------------------------------
\begin{figure}[t]
\includegraphics[width=3in]{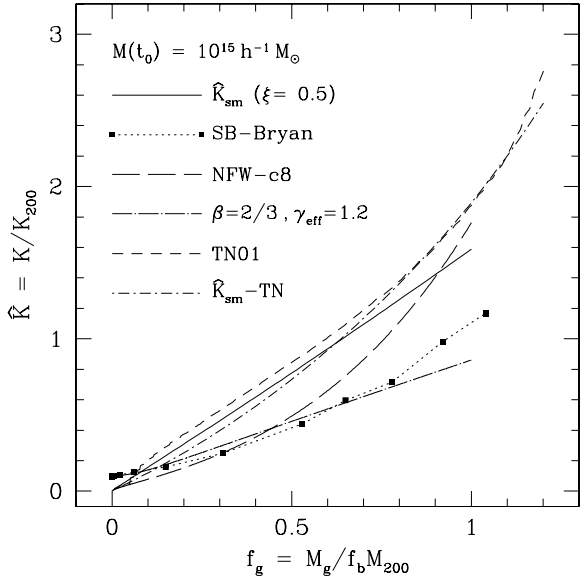}
\caption{ \footnotesize
Dimensionless entropy profiles $\hat{K}(f_g)$ showing the
fractional amount of gas $f_g = M_g / f_b M_{200}$ with entropy
$< \hat{K}$. Solid lines
show the entropy profile derived from equation~(\ref{eq-khatcold})
using the accretion history for a $10^{15} \, h^{-1} \, M_\odot$
cluster from \S~\ref{sec-acchist}.  Filled squares and the
dotted line show the profile from the Bryan ``Santa Barbara''
cluster (SB-Bryan).  The long-dashed line depicts the 
unmodified profile for an NFW halo with concentration $c=8$
from Voit et al. (2002).  The dot-dashed line with long
dashes illustrates the $\beta$-model approximation to
observations described in the text.  The short-dashed line shows the
profile for a $10^{15} \, h^{-1} \, M_\odot$ cluster computed
by Tozzi \& Norman (2001), and the dot-dashed line with short
dashes shows the result of equation~(\ref{eq-khatcold}) 
for the TN01 accretion history in Figure~\ref{mtfits}. 
\label{ent_comp_all_1}}
\end{figure}
% ----------------------------------------

% ----------------------------------------
\begin{figure}[t]
\includegraphics[width=3in]{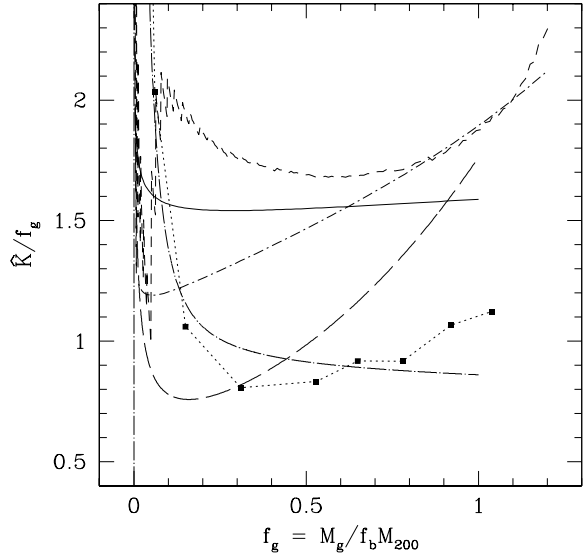}
\caption{ \footnotesize
Normalized entropy profile $\hat{K}/f_g$ versus $f_g$.  
Lines have the same meanings as in Figure~\ref{ent_comp_all_1}.
Note that the entropy profiles based on simulations and simulations
generally follow $\hat{K}/f_g \approx 1$, while those based on
smooth accretion follow $\hat{K}/f_g \approx 1.5-2$.
\label{ent_comp_all_2}}
\end{figure}
% ----------------------------------------

Figure~\ref{ent_comp_all_1} compares the dimensionless entropy
profile $\hat{K}_{\rm sm}(f_g)$ from the smooth accretion model, 
assuming $\xi = 0.5$ for a $10^{15} \, h^{-1} \, M_\odot$ cluster, 
with several other entropy profiles computed in different ways.
The filled squares connected by a dotted line show the entropy
profile from a numerically simulated cluster, the ``Santa Barbara''
cluster (Frenk et al. 1999) created by an adaptive mesh refinement 
code (Norman \& Bryan 1998, Bryan 1999).  The dark-matter density
profile of this cluster closely follows the NFW form 
$\rho_{\rm dm} \propto [r(1+cr/r_{200})^2]^{-1}$ 
(Navarro, Frenk, \& White 1997) with concentration $c=8$,
and the gas density also follows this form at $r \gtrsim 0.1 r_{200}$.
Thus, we also show the entropy profile of gas in hydrostatic
equilibrium in an NFW potential with $c=8$ when the gas density
is precisely proportional to the dark-matter density (long-dashed
line; NFW-8).  This is the prescription sometimes used to compute
the form of the entropy profile before it is modified by non-gravitational
processes (e.g., Bryan 2000; Voit \& Bryan 2001; Wu \& Xue 2002a,b;
Voit et al. 2002).   It underpredicts the core entropy found by
simulations, possibly because it does not account for energy transfer from
the dark matter to the baryons (e.g., Navarro \& White 1993),
and it overpredicts the entropy in the outer regions, where the
cluster is not in hydrostatic equilibrium. In order to mimic 
the observed entropy profiles of clusters, we use a $\beta$-model
density distribution with $\beta = 2/3$ and core radius of
$0.1 r_{200}$ (Cavaliere \& Fusco-Femiano 1976) with a polytropic
relation ($T \propto \rho^{\gamma_{\rm eff} -1}$, 
$\gamma_{\rm eff} = 1.2$) relating density to temperature.  
We normalize the temperature so that $T = T_{200}$ at the 
core radius, which implies $T \sim 0.5 T_{200}$ at $r_{200}$.
Note that this empirically derived entropy profile
closely corresponds to the numerically simulated profile
without any tuning of $\beta$ or $\gamma_{\rm eff}$.

The nearly linear slope of the entropy profile 
from the smooth accretion model is similar to that 
of the simulated cluster, but its normalization
is too large.  Outside the core of the simulated
cluster, we find $\hat{K} \approx f_g$,  while the smooth 
accretion model with $\xi = 0.5$ yields $\hat{K} \approx 1.6 f_g$.  
Dimensionless entropy in the model of Tozzi \& Norman (2001) 
has an even higher normalization, $\hat{K} \approx 1.8 f_g$ 
for a $10^{15} \, h^{-1} \, M_\odot$ cluster, 
as indicated by the short-dashed line (see 
Figure~\ref{ent_comp_all_2}).  Much of the difference 
in normalization between their numerical model and the 
analytical model developed in this
paper stems from the different accretion history.  Their accretion
law converts to $M(t) \propto t^{0.931 - \log t/t_0}$, as shown by the
dotted line in Figure~\ref{mtfits}.  Plugging this accretion law
into our analytical model with $\xi = 0.45$, as suggested by 
the analysis of \S~\ref{sec-preheat} and Appendix~A, produces much 
better agreement at $f_g > 0.5$, shown by the dot-dashed line 
($\hat{K}_{\rm sm}$-TN) in Figures~\ref{ent_comp_all_1} and
\ref{ent_comp_all_2}. The remaining discrepancy at $f_g \lesssim 0.5$ 
comes mostly from the preheating assumed by Tozzi \& Norman (2001), 
as we will discuss in \S~\ref{sec-preheat}.

% ----------------------------------------
\begin{figure}[t]
\includegraphics[width=3in]{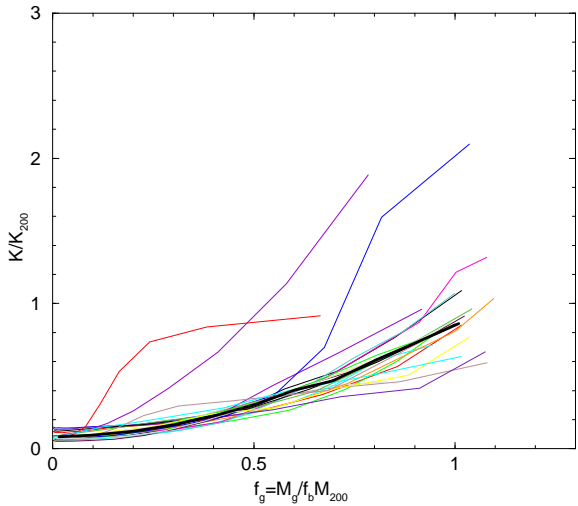}
\caption{ \footnotesize
Dimensionless entropy profiles $K(M_g)/K_{200}$ for
twenty-four halos from a numerical simulation
without radiative cooling or feedback.  The objects were
drawn from simulation L50+ described in Bryan \& Voit (2001).
They span a mass range from $2.5 \times 10^{13} \, h^{-1} \, 
M_\odot$ to $3.5 \times 10^{14} \, h^{-1} \, M_\odot$, 
yet they generally have nearly linear entropy 
distributions with the same normalization found in
more massive clusters, $K(r_{200}) \approx K_{200}$,
with no significant trend in mass.
A bold solid line shows the median profile.
\label{L50_nf}}
\end{figure}
% ----------------------------------------

This discrepancy between the entropy generated by smooth accretion
and the entropy produced in simulations is even larger in groups
because of their slower accretion rates.  The logarithmic 
derivative $d \ln \eta / d \ln t$ for groups is only about half 
the large-cluster value (see Figure~\ref{mtfits}).
According to equation (\ref{eq-khatcold}), the dimensionless
entropy $\hat{K}_{\rm sm}$ should correspondingly be $\sim$50\% larger,
approaching $\sim 3 K_{200}$ at $f_g \sim 1$.  
Yet, the dimensionless entropy profiles of simulated groups remain
similar to those of simulated clusters.  Figure~\ref{L50_nf}
shows entropy distributions for the twenty-four highest-mass objects
in $\Lambda$CDM simulation L50+ of Bryan \& Voit (2001).  
Except for a few outliers, these halos share a dimensionless 
entropy distribution similar to that of the more massive
Santa-Barbara cluster, even though their masses range from
$2.5 \times 10^{13} \, h^{-1} \, M_\odot$ to $3.5 \times 10^{14} 
\, h^{-1} \, M_\odot$.

\subsection{Departures from Smooth Accretion}
\label{sec-depsmooth}

The most obvious explanation for the difference in normalization
between the smooth accretion model and those derived from
simulations and observations is that the accreting gas in a 
more realistic model would be lumpy, not smooth.  
Any inhomogeneity in the density of gas accreting onto a 
cluster tends to reduce the mean post-accretion entropy, 
as long as the velocity of the accreting gas remains unchanged.  
The reduction occurs because the postshock entropy scales
as $v_{\rm ac}^2 \rho_1^{-2/3}$ and the mass-weighted
mean value of $\rho_1$ is larger if there is any inhomogeneity,
anisotropy, or unsteadiness in the accretion flow
(see Appendix~B).

This sensitivity of the postshock entropy to the density
distribution of incoming material means that the entropy
normalization of a cluster or group reflects the lumpiness
of the gas that accreted onto it.  In Section~\ref{sec-evidence} 
we will show that this effect may actually be important---the
elevated normalization of entropy in groups suggests that
accretion of baryons onto groups was smoother than
accretion of baryons onto clusters.
However, we will first continue with our analysis of smooth
accretion in order to include preheating and cooling in the
analytical model.

\section{Smooth Accretion of Preheated Gas}
\label{sec-smoothpre}

Many authors have argued that preheating of gas that
accretes onto a cluster is needed to explain the observed
slope of the $L$--$T$ relation (e.g., Kaiser 1991; Evrard \&
Henry 1991; Navarro, Frenk, \& White 1995; Cavaliere, Menci,
\& Tozzi 1997; Balogh \etal 1999; Ponman \etal 1999).  Depending 
on how the analysis is done, the necessary amount of preheating
ranges from $T n_e^{-2/3} \sim 100 \, \keV \, {\rm cm}^2$
to over $400 \, \keV \, {\rm cm}^2$, corresponding to
$K \sim 10^{33} \, {\rm erg \, cm^2 \, g^{-5/3}}$ 
to $4 \times 10^{33} \, {\rm erg \, cm^2 \, g^{-5/3}}$
(but see \S~\ref{sec-polyfloor}).
If the preshock entropy level $K_1$ is comparable to $K_{\rm sm}$,
then we can no longer assume that the accreting gas is
pressureless.  In particular, the Mach number of the
accretion shock is $\stackrel {\propto} {\sim} (K_{\rm sm} 
/ K_1)^{1/2}$, meaning that production of postshock
entropy depends on the level of preheating.  

In this section we derive an entropy jump condition that accounts
for preheating and outline how preheating affects the position
of the shock radius, suppressing the postshock entropy.
Because radiative cooling can offset some of the effects
of preheating, we also derive a simple approximation to
treat cooling.  Then, we add both preheating and cooling
to our simple analytical entropy profiles and compare them
with the numerically modeled entropy profiles of Tozzi \& 
Norman (2001).  Because the analytical profiles closely
match the numerically computed ones, we conclude that our
analytical model is a good representation of smooth accretion.

\subsection{Entropy Jump Condition}
\label{sec-entjump}

A jump condition for entropy production when preheated gas passes 
through a shock can be derived from the jump conditions for other
quantities.  From the density jump condition, we get
\begin{equation}
  v_{\rm in} = v_1 - v_2 
      = \frac {3} {4} \left( 1 - \frac {1} {{\cal M}^2} \right) v_1
      \; \; ,
\end{equation}
where $v_1$ and $v_2$ are the preshock and postshock gas velocities,
respectively, in the rest frame of the shock, and $v_{\rm in}$ is 
the velocity of incoming gas relative to the postshock gas.
The Mach number ${\cal M} = (3 \rho_1 v_1 / 5 P_1)^{1/2}$ 
of the shock is then determined by
\begin{equation}
  \frac {({\cal M}^2 - 1)^2} {{\cal M}^2} 
        = \frac {16} {15} \frac {\rho_1 v_{\rm in}^2} {P_1} 
          \; \; .
\end{equation}
Noting that $K_1 = P_1 \rho_1^{-5/3}$ and setting $v_{\rm in} = 
v_{\rm ac}$, we obtain
\begin{equation}
  \frac {({\cal M}^2 - 1)^2} {{\cal M}^2} 
   = \frac {4^{8/3}} {5} \frac {K_{\rm sm}} {K_1}
          \; \; .
\end{equation}
Solving this quadratic equation for the larger root gives
\begin{equation}
 {\cal M}^2 = \frac {4^{8/3}} {5} \frac {K_{\rm sm}} {K_1}
              \left( \frac {1 + q_K + \sqrt{1+ 2 q_K}} {2} \right)
                   \; \; ,
\end{equation}
where $q_K \equiv 10 K_1 / 4^{8/3} K_{\rm sm}$.

% ----------------------------------------
\begin{figure}[t]
\includegraphics[width=3in]{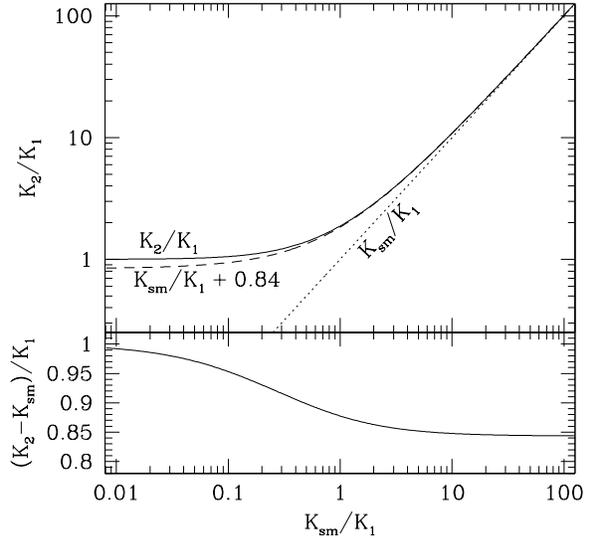}
\caption{ \footnotesize
Modification of postshock entropy owing to preheating.  
Accretion of cold gas produces a postshock entropy $K_{\rm sm}
= v_{\rm ac}^2 / 3 (4 \rho_1)^{2/3}$ in the high 
Mach-number limit.  Preheating that raises the entropy 
of incoming gas to a preshock value $K_1$ 
weakens the shock but raises the postshock entropy $K_2$ 
above $K_{\rm sm}$ by an amount $\approx 0.84 K_1$.  
The solid line in the upper panel shows $K_2/K_1$ as given by
equation (\ref{eq-kjump}).  The dotted line shows the postshock
entropy $K_{\rm sm}/K_1$ one finds if preheating is ignored.  
The dashed line shows an approximate correction for preheating:
$K_2 \approx K_{\rm sm} + 0.84 K_1$.  The lower panel more precisely
shows the difference between $K_2$ and $K_{\rm sm}$.
\label{kjump}}
\end{figure}
% ----------------------------------------

Now we can express the jump conditions in terms of $K_{\rm sm}/K_1$,
\begin{eqnarray}
  \frac {P_2} {P_1} & = & 4^{5/3} \frac {K_{\rm sm}} {K_1} 
                 \left( \frac {1 + \sqrt{1+2q_K}} {2} +
                        \frac {2q_K} {5} \right)  \nonumber \\
  \frac {\rho_1} {\rho_2} & = & \frac {1} {4} 
                 \left( 1 + \frac {3q_K} {1+q_K+\sqrt{1+2q_K}} \right)
                       \; \; ,  \nonumber
\end{eqnarray}
which lead to
\begin{eqnarray}
  K_2 & = & K_{\rm sm} 
               \left( \frac {1+\sqrt{1+2q_K}} {2} +
                        \frac {2q_K} {5} \right)  \; \; \; \nonumber \\
    ~ & ~ &  \; \; \times  
               \left( 1 + \frac {3q_K} {1+q_K+\sqrt{1+2q_K}} \right)^{5/3}
       \; \; .
       \label{eq-kjump}
\end{eqnarray}
In the limit $q_K \ll 1$, equivalent to $K_{\rm sm}/K_1 \gg 0.25$,
this expression reduces to $K_2 \approx K_{\rm sm} + 0.84 K_1$
(see Figure~\ref{kjump}; Dos Santos \& Dor\'e (2002) arrive at a 
similar approximation following a different route.)
From this point of view, simply adding a constant value to the
entropy profile produced without preheating, as in the shifted 
models of Voit \etal (2002), seems like a good representation 
of the effects of preheating.

\subsection{Preheating and the Shock Radius}
\label{sec-preheat}

If the entropy of accreting gas substantially exceeds that
which would be created by a shock, then we can no longer
assume $\xi \approx 0.5$.  Because the pressure just inside
the shock front must balance the ram pressure of accreting
gas, the position of the shock radius depends on both the 
current accretion rate and the internal structure of the
cluster.  When extra entropy, over and above that produced 
by cold accretion, is present within the cluster, the equilibrium
shock radius is larger.  If the extra entropy substantially
exceeds $K_{\rm sm}$, then the equilibrium radius becomes
so large that shock heating is essentially turned off.

\begin{figure}[t]
\includegraphics[width=3in]{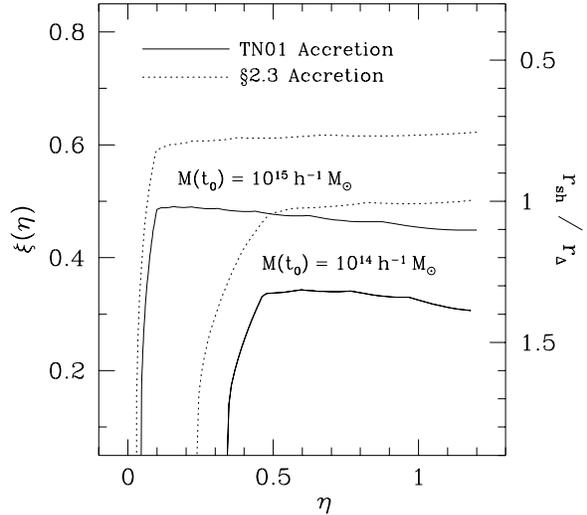}
\caption{ \footnotesize
Shock-radius parameter $\xi$ as a function of the
accreted mass $\eta(t) = M(t)/M_{200}$.  The solid lines
show solutions for $\xi$ using the model developed in
Appendix~A for the cluster parameters of Tozzi \& Norman
(2001).  The dotted lines show solutions found using the
accretion histories of \S~\ref{sec-acchist} but otherwise
identical cluster parameters.  Lines at the upper left show
clusters of mass $10^{15} \, h^{-1} \, M_\odot$; those at the
lower right show clusters of mass $10^{14} \, h^{-1} \, M_\odot$.
In all cases, the initial shock radius greatly exceeds the virial 
radius ($r_\Delta = r_{\rm ta} / 2$) because the entropy of preheating 
($K_1 = 3 \times 10^{33} \, {\rm erg \, cm^2 \, g^{-5/3}}$) 
grossly distends the intracluster medium.  After the cluster's
gravity overcomes the entropy of preheating, the equilbrium value
of $\xi$ is determined primarily by the accretion rate onto 
the cluster.
\label{xi_tn}}
\end{figure}

Appendix~A presents an approximate analytical 
solution for the equilibrium shock radius in terms of the 
associated shock-velocity parameter $\xi$ that applies 
when preheated gas smoothly accretes onto a cluster.  
Figure~\ref{xi_tn} shows the relation between $\xi$ and
$M(t)$ arising from the accretion histories and concentration
parameters used by Tozzi \& Norman (2001).  As in the
numerical models of Tozzi \& Norman (2001), the initial
shock radius greatly exceeds the virial radius ($r_\Delta
= r_{\rm ta} / 2$) because the
entropy of preheating, $K_1 = 3 \times 10^{33} \, {\rm erg \,
cm^2 \, g^{-5/3}}$, is much larger than the characteristic
entropy of the nascent dark-matter halo.  Eventually,
the halo's gravity overcomes the effects of preheating,
pulling the shock radius inward to a stable value of
$\xi$.  The more massive cluster ($10^{15} \, h^{-1} \,
M_\odot$) achieves a larger value of $\xi \approx 0.45$ 
primarily because its faster accretion rate is more 
effective at compressing the gas internal to the shock 
front.  In the less massive cluster ($10^{14} \, h^{-1} \,
M_\odot$), accretion is more gradual, resulting in less
compression and a smaller value of $\xi \approx 0.35$.
For comparison, we also show solutions for $\xi$ using the
accretion histories of \S~\ref{sec-acchist} but otherwise
identical cluster parameters.  In both cases, faster
accretion produces a higher value of $\xi$.

% ----------------------------------------
% ----------------------------------------

The constancy of $\xi$ in the shock-dominated regimes of these 
clusters is not trivial.  On one hand, the rise in halo concentration
with time tends to pull the shock radius slightly inward.  On the
other hand, the decrease in accretion rate with time allows the
shock radius to slowly increase.  In these particular models,
the two tendencies nearly cancel.  Note also that the model we
are using is not a good representation for $\xi \ll 0.5$ because
we have assumed that the time at which infalling gas reaches 
the shock radius is twice the time it took to reach its turnaround
radius.

\subsection{Correcting for Cooling}
\label{sec-cooling}

\begin{figure}[t]
\includegraphics[width=3in]{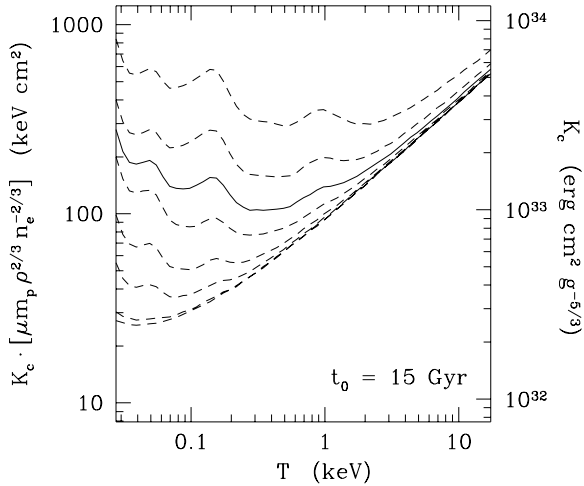}
\caption{ \footnotesize
Threshold entropy for cooling within $t_0 = 15$~Gyr as a function 
of temperature and metallicity.  The quantity $K_c(T)$ is defined
in equation~(\ref{eq-kc}).  Reading down from the top, the lines
show $K_c(T)$ for metallicities of $\log Z/Z_\odot = 0.5$, 0.0, -0.5,
-1.0, -1.5, -2.0, 3.0, and zero metallicity, based on the cooling functions 
of Sutherland \& Dopita (1993).  The single solid line highlights 
the value of $K_c(T)$ for a typical cluster metallicity 
of $\log Z/Z_\odot = -0.5$. 
\label{coolents}}
\end{figure}

Radiative cooling can significantly modify the entropy distribution
near the center of a cluster, if the entropy of preheating is
not excessively large.  One illustrative way to express the 
effects of radiative losses is with the following equation for 
the resulting change in entropy:
\begin{equation}
  \frac {d K^{3/2}} {dt} = - \frac {3} {2} \frac {K_c^{3/2}(T)} {t_0} \; ,
\end{equation}
where
\begin{equation}
  K_c(T) = \left[ \frac {2} {3} 
                        \left( \frac {n_e n_p} {\rho^2} \right)
                        \frac {T^{1/2} \Lambda(T)} {(\mu m_p)^{1/2}}
                          \right]^{2/3} t_0^{2/3}
\label{eq-kc}
\end{equation}
is the entropy level at which constant-density gas at temperature $T$ 
radiates an energy equivalent to its thermal energy in the time $t_0$ and
$\Lambda(T)$ is the standard cooling function.  Figure~\ref{coolents}
shows how $K_c(T)$ depends on $T$ in the models of Sutherland \& Dopita
(1993) for various metallicities.  The solid line highlights $K_c(T)$ 
for a typical cluster metallicity of $\log Z/Z_\odot = -0.5$.

% ----------------------------------------
% ----------------------------------------

In a simple spherical accretion model, a gas shell experiences no
additional heat input after it has passed through a shock front.
Gravitational compression can raise the temperature of the shell,
but its entropy does not rise.  Thus, radiative cooling inexorably
lowers the entropy of the shell, resulting in
\begin{equation}
 K_{\rm mod}(\eta) = \left[ K_2^{3/2}(\eta) 
                      - K_{\rm rad}^{3/2}(\eta)  
                           \right]^{2/3} \; \; ,
\end{equation}
where
\begin{equation}
 K_{\rm rad}(\eta) \equiv \left[ \frac {3} {2t_0} 
                 \int_{t(\eta)}^{t_0} K_c^{3/2} (T) \, dt 
                      \right]^{2/3}  \; \; .
\end{equation}
The term $K_{\rm rad}(\eta)$ characterizes the entropy lost
by the gas shell that accretes at time $t(\eta)$ after it
passes through the accretion shock.  In the comparisons that
follow, we approximate $K_{\rm rad}(\eta)$ by setting
$T$ equal to $T_\phi = G M \mu m_p/ r_{\rm ta}$, the characteristic 
temperature of the dark matter halo at time $t$, which
directly ties the integrand to the accretion history.\footnote{Note 
that this definition of characteristic
temperature differs from $T_{200}$ by a small amount depending
on the overdensity parameter $\Delta$ and the halo concentration; 
for $\Omat = 0.3$ and halo concentration $c = 8$ it equals
$\approx 0.9 T_{200}$ at the present time.}   
We also assume a constant metallicity of $\log Z / Z_\odot 
= -0.5$.

\subsection{Comparisons with Numerical Models}

Having developed expressions for how preheating and
radiative cooling affect the postshock entropy of smooth 
accreting gas, we are now in a position to compare our
analytical accretion model with the numerical models 
of Tozzi \& Norman (2001).  Their standard models
assumed an initial entropy $K_1 = 3 \times 10^{33} \, 
{\rm erg \, cm^2 \, g^{-5/3}}$ and a constant
metallicity of $0.3 Z_\odot$.  The underlying cosmology
was $\Lambda$CDM with $\Omat = 0.3$, $\Olam = 0.7$, and
$h = 0.65$, so that $f_b = 0.158$ and $t_0 = 14.5$~Gyr.

% ----------------------------------------
\begin{figure}[t]
\includegraphics[width=3in]{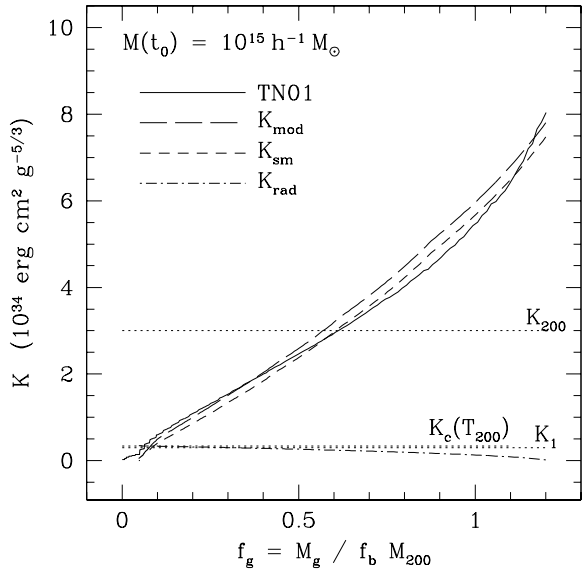}
\caption{ \footnotesize
Entropy profiles for preheated gas that has accreted into a 
cluster with $M(t_0) = 10^{15} \, h^{-1} \, M_\odot$.  The
solid line shows the numerical model of Tozzi \& Norman (2001).
The long-dashed line shows the analytical model ($K_{\rm mod}$)
from this paper, computed from the accretion history assumed 
by Tozzi \& Norman (2001), including corrections for preheating
and radiative cooling.  The short-dashed line indicates the
entropy profile ($K_{\rm sm}$) that results if these corrections
are not applied.  Dotted lines show $K_1$, the initial entropy
of preheating, $K_c(T_{200})$ the cooling threshold of the
final cluster, and the halo entropy scale $K_{200}$.  
The dot-dashed line shows $K_{\rm rad}$, the correction
term for radiative cooling.  Because both $K_1$ and $K_{\rm rad}$
are small compared to $K_{200}$, these corrections have 
only a minor effect on the global entropy profile outside 
the very center.
\label{ent_comp_tn15}}
\end{figure}
% ----------------------------------------

Figure~\ref{ent_comp_tn15} shows the comparison for a
$10^{15} \, h^{-1} \, M_\odot$ cluster.  In this case,
the analytical model closely tracks the nearly linear
behavior of the numerically computed profile at $f_g >
0.1$.  Outside the lowest-entropy regions, the corrections
owing to preheating and radiative cooling are relatively
small because $K_1$ and $K_{\rm rad}$ are both less than
10\% of the cluster's characteristic entropy.  However,
preheating is important early in the accretion history,
when the characteristic entropy of the young halo is much 
smaller, and the relationship between preheating and
cooling has important implications for the fate of the
central gas.  

Figure~\ref{xi_tn} shows that the first $\sim 5$\% 
of the cluster's gas accretes nearly isentropically
because $\xi \ll 0.5$, so its entropy remains at the initial
value of $K_1 = 3 \times 10^{33} \, {\rm erg \, cm^2 \, g^{-5/3}}$. 
After this gas accretes, it is subject to radiative cooling. 
The approximate cooling model of \S~\ref{sec-cooling} gives
$K_{\rm rad} \approx 3.5 \times 10^{33} \, {\rm erg \, cm^2 \,
g^{-5/3}}$ for the earliest gas to accrete, $\sim 15$\% higher
than the entropy of preheating.  (Note that the cooling
threshold $K_c(T_{200})$ of cluster gas at time $t_0$
closely corresponds to the integral for $K_{\rm rad}$.)
The entropy of the innermost gas therefore drops to zero, 
implying a condensed fraction $\sim 5$\%, which is ten 
times greater than the $\sim 0.5$\% condensed fraction in 
the Tozzi \& Norman (2001) model.  Yet, if the cooling rate 
in the analytical model were $\sim 20$\% smaller, or  
preheating were $\sim 20$\% greater, the 
condensed fraction would be zero.  This underscores
the importance of accounting for entropy produced by
internal feedback following cooling and condensation of 
intracluster gas (e.g., Voit \& Bryan 2001).

% ----------------------------------------
\begin{figure}[t]
\includegraphics[width=3in]{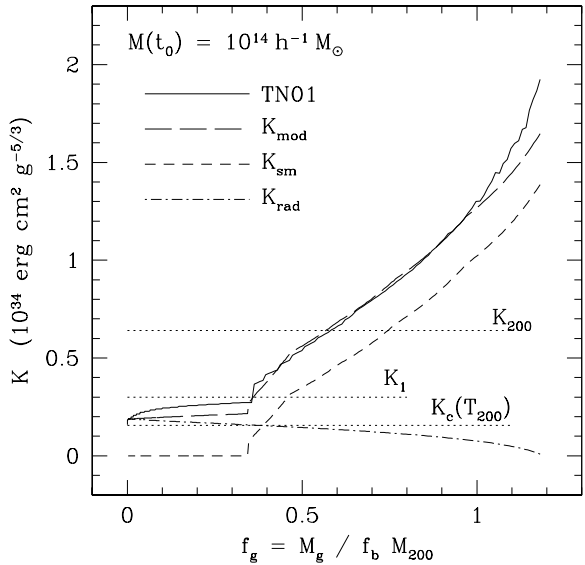}
\caption{ \footnotesize
Entropy profiles for preheated gas that has accreted
into a cluster with $M(t_0) = 10^{14} \, h^{-1} \, M_\odot$.  Lines
have the same meanings as in Figure~\ref{ent_comp_tn15}.  Because the
entropy of preheating is relatively significant, the accretion radius
is initially much larger than the virial radius (see Figure~\ref{xi_tn}),
and the first third of the cluster's gas accretes isentropically.
After a strong shock develops, the entropy profile exceeds $K_{\rm sm}$
by $\approx 0.84 K_1$.  Because $K_{\rm rad}$ is not much less than 
$K_1$, radiative cooling significantly diminishes the central entropy.
\label{ent_comp_tn14}}
\end{figure}
% ----------------------------------------

Preheating has a much more significant impact on
the $10^{14} \, h^{-1} \, M_\odot$ cluster, yet the analytical
model remains a good match to the numerical model.  In this case,
over 30\% of the gas accretes isentropically, while $\xi \ll
0.5$ (see Figure~\ref{xi_tn}).  Even after the accretion
shock strengthens, there is still a significant constant
offset ($\approx 0.84 K_1$) between $K_{\rm sm}$ and
the Tozzi \& Norman (2001) model.  Cooling diminishes the
final entropy of the gas that accreted isentropically,
and this decrease is again slightly larger in the analytical
model than in the numerical model.  Note also that $K_c(T_{200})$
remains a good approximation to $K_{\rm rad}$ for the central
gas.

% ----------------------------------------
\begin{figure}[t]
\includegraphics[width=3in]{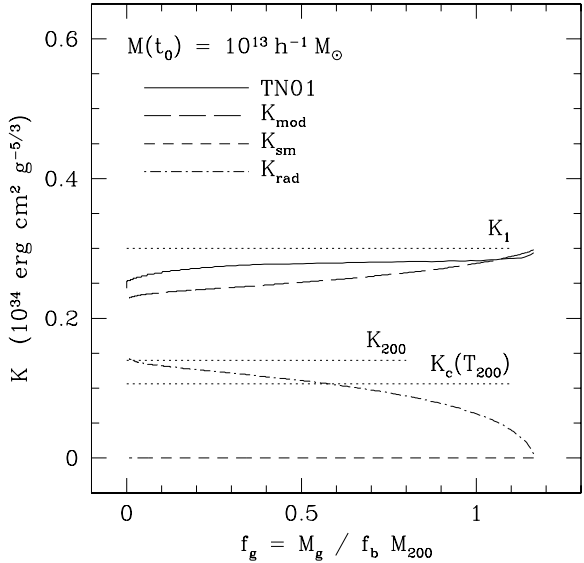}
\caption{ \footnotesize
Entropy profiles for preheated gas that has accreted
into a cluster with $M(t_0) = 10^{13} \, h^{-1} \, M_\odot$.  Lines
have the same meanings as in Figure~\ref{ent_comp_tn15}.  In this
case, the entropy of preheating exceeds the characteristic entropy
of the final cluster, so the accretion radius never approaches the
virial radius.  Accretion is therefore virtually isentropic.
As in the $10^{13} \, h^{-1} \, M_\odot$ cluster, radiative
cooling significantly lowers the central entropy.
\label{ent_comp_tn13}}
\end{figure}
% ----------------------------------------

The characteristic entropy of a $10^{13} \, h^{-1} \, M_\odot$ halo
is so small that $3 \times 10^{33} \, {\rm erg \, cm^2 \, g^{-5/3}}$
of preheating dominates everything.  The shock radius never approaches 
the virial radius, so all the gas accretes isentropically.  
Because $K_{\rm rad} \approx 0.5 K_1$ for the earliest gas to accrete, 
cooling diminishes the central entropy by $\sim 25$\%. Again, 
the analytical model calls for slightly more cooling than the 
numerical model.

In all three cases, our approximate analytical model for smooth 
accretion seems to represent with reasonable accuracy all the 
important processes operating in the numerical models.  The analytical
models show that preheating can suppress radiative cooling in 
a given mass shell if $K_{\rm sm} + 0.84 K_1$ exceeds $K_{\rm rad} 
\approx K_c(T_{200})$.  However, the cores of these objects
become isentropic if $K_1$ is comparable to $K_{200}$ and larger
than $K_{\rm rad}$ (see Figures~\ref{ent_comp_tn14} and 
\ref{ent_comp_tn13}).

\subsection{Inhomogeneous Preheated Gas}
\label{sec-inhomopre}

What effect does inhomogeneity have on accretion of preheated
gas?  According to \S ~\ref{sec-entjump}, the postshock entropy
of preheated gas is
\begin{equation}
 K_2 \approx \frac {v_{\rm in}^2} {3 (4 \rho_1)^{2/3}} + 0.84 K_1
\label{eq-k2approx}
\end{equation}
As long as the shock velocities $v_{\rm in}$ are comparable
to the accretion velocity $v_{\rm ac}$, the postshock entropy
will depend primarily on the preshock density $\rho_1$ and the
preshock entropy $K_1$.  When we consider how preheating
affects the case of lumpy, hierarchical accretion, the second term
containing $K_1$ is relatively straightforward to apply and
produces a correction no greater than the entropy of preheating.
However, if preheating can significantly smooth the accreting gas,
substantially lowering the mean mass-weighted preshock density, 
then the increase in postshock entropy owing to the $v_{\rm in}^2 /
3 (4 \rho_1)^{2/3}$ term can be considerably larger than $K_1$.  
We will consider some observational evidence for this effect 
in section~\ref{sec-evidence}.

Notice also that equation (\ref{eq-k2approx}) does not
explicitly depend on the temperature of the preheated gas.
We are choosing to focus on entropy and density, rather 
than on preshock temperature, because this way of 
casting the problem implicitly accounts
for preshock heating owing to adiabatic compression.
In fact, focusing on preshock temperature can be misleading
because adiabatic heating actually {\em decreases} 
the postshock entropy of an accreting gas blob;  compression 
increases the preshock density, lowering $K_{\rm sm}$
without changing $K_1$.  The entropy gain across the
shock is smaller because the preshock temperature is
larger than it would have been without the compression,
so the heat input released by thermalization of the
incoming kinetic energy produces less entropy.
Thus, the postshock entropy is more closely related 
to the preshock density distribution than to the 
preshock temperature distribution.

\section{Evidence for Smoothed Accretion}
\label{sec-evidence}

The results of the previous two sections imply that 
the impact of preheating on lumpy accretion qualitatively
differs from its impact on smooth accretion.  Entropy input 
preceding smooth accretion suppresses entropy 
production at the accretion shock because it pushes the 
accretion radius beyond the virial radius (e.g., Balogh
\etal 1999; Tozzi \& Norman 2001; \S~\ref{sec-preheat}).  
If the entropy of preheating exceeds the characteristic 
entropy of the halo, then smooth accretion becomes 
virtually adiabatic, leading to a nearly isentropic 
entropy distribution.  However, dense lumps of accreting
gas will not necessarily be prematurely halted by the distended
intracluster medium, so it is not clear that entropy production 
by lumpy accretion will be diminished in quite the same way.

On the contrary, preheating might actually {\em enhance}
entropy generation over what would normally be produced in
the lumpy accretion mode.  Because preheating tends to lower 
the density of gas within small halos, it smooths the density 
distribution accreting onto larger halos, increasing 
the efficiency with which accretion shocks generate entropy.  
If preheating ejects a majority of the gas from small 
halos, then this effect might be strong enough to 
cause a transition from lumpy accretion to smooth 
accretion on the mass scale of groups.  In that case, 
smooth accretion onto groups would produce an entropy 
normalization $\sim$2-3 times higher than that expected 
from self-similar scaling of clusters.  Furthermore, if
the initial level of preheating ($K_1$) is relatively
small compared to the characteristic entropy of the
final halo ($K_{200}$), then the final halo need not
have a substantial isentropic core.

This section examines some intriguing pieces of observational 
evidence suggesting that preheating alters the entropy
distributions of groups primarily by smoothing the gas
that accretes onto them.  Thus, the large amounts of preheating 
needed to explain the $L$-$T$ relation in the case of
isentropic groups might be unnecessary.  First, we show that groups with 
flat entropy gradients cannot simultaneously be consistent 
with both the observed $L$-$T$ relation 
and observations of core entropy.  Polytropic models of
groups with an effective adiabatic index $\gamma_{\rm eff} 
\approx 1.2$ are more consistent with these constraints,
indicating that the entropy gradients of groups are 
similar to those of clusters.\footnote{Direct observations
of intragroup entropy that appeared after this paper was
submitted support this conclusion (Mushotzky \etal 2003;
Pratt \& Arnaud 2003).}  Then, we show that these polytropic 
models imply entropy levels at the virial radii of 
groups that exceed the maximum value of $K(r_{200}) \approx 
K_{200}$ expected from lumpy accretion, 
a conclusion supported by recent measurements of entropy
at the outskirts of groups (Finoguenov \etal 2002).
The excess entropy implied by observations is similar 
to the amount expected from smooth accretion onto groups,
perhaps indicating a transition from lumpy
accretion to smooth accretion at a mass
scale $\sim 10^{14} \, h^{-1} \, M_\odot$
owing to preheating in smaller halos.

\subsection{Entropy Gradients of Groups}
\label{sec-polyfloor}

The relationship between the X-ray luminosities and
temperatures of groups and clusters has long been
assumed to indicate an early episode of preheating, but the
necessary amount of preheating is a matter of
some debate.  Purely gravitational structure formation
calls for $L \propto T^2$ (e.g., Kaiser 1986), yet observations
show $L \propto T^{2.5-3}$ (e.g., Edge \& Stewart 1991).
This steepening of the $L$-$T$ slope requires
some sort of non-gravitational modification of the core 
entropy (Kaiser 1991; Evrard \& Henry 1991; Bower 1997; 
Voit \etal 2002).  

Some models are able to fit this relation with a core
entropy $\sim 100$-$150 \, {\rm keV \, cm^2}$ $(\sim 1.0-1.5
\times 10^{33} \, {\rm erg \, cm^3 \, g^{-5/3}}$) (e.g.,
Cavaliere \etal 1997, 1998, 1999; Voit \& Bryan 2001; 
Bialek, Evrard, \& Mohr 2001; dos Santos 
\& Dor\'e 2002).  Other models require $\sim 200$-$400 
\, {\rm keV \, cm^2}$ $(\sim 2.0$-$4.0 \times 10^{33} \, 
{\rm erg \, cm^3 \, g^{-5/3}}$) (e.g., Balogh \etal 1999;
Tozzi \& Norman 2001; Babul \etal 2002).  The primary
difference between these two classes of models is
that those requiring large amounts of preheating
have isentropic entropy profiles extending to a large fraction
of the virial radius, while those requiring less central
entropy assume a significant entropy gradient throughout
the group.

We can quantify this difference using the polytropic
models described by equations (\ref{eq-tg}) through (\ref{eq-x}) 
of Appendix~A.  These models require
two boundary conditions and an effective polytropic
index $\gamma_{\rm eff}$.  For this application, we
set one boundary condition so that the pressure at
$r_{200}$ is the same for all
models, and we choose that pressure to be equal to
the $\gamma_{\rm eff} = 1.2$ case with a total
gas mass $f_b M_{200}$ within $r_{200}$.  For the other 
boundary condition we fix the entropy at at $0.1r_{200}$. 
Each model is therefore determined by 
the halo mass $M_{200}$, the halo concentration parameter
$c_{200}$, the polytropic index $\gamma_{\rm eff}$,
and the core entropy $K_{0.1} \equiv K(0.1r_{200})$.
For each halo mass, we consider a set of polytropic
models with $\gamma_{\rm eff} = 1.2$ and a set of
isentropic models ($\gamma_{\rm eff} = 5/3$).

\begin{figure}[t]
\includegraphics[width=3in]{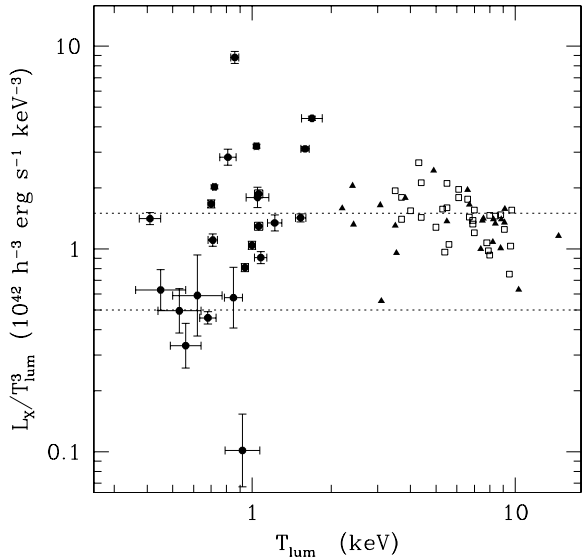}
\caption{ \footnotesize
Luminosity-temperature relation for groups and clusters
in terms of $L_X/T_{\rm lum}^3$.  Solid triangles 
show measurements of clusters with insignificant cooling flows
compiled by Arnaud \& Evrard (1999).  Open squares show
cooling-flow corrected measurements by Markevitch (1998).
Solid circles show group data from Helsdon \& Ponman (2000).
Note that the quantity $L_X/T_{\rm lum}^3$ remains roughly
constant at $\sim 1.5 \times 10^{42} \, h^{-3} \, {\rm erg 
\, s^{-1} \, keV^{-3}}$ (upper dotted line), dipping to 
slightly lower levels ($\sim 0.5 \times 10^{42} \, h^{-3} 
\, {\rm erg \, s^{-1} \, keV^{-3}}$ (lower dotted line)
below $\sim 1$~keV.
\label{lt3rel}}
\end{figure}

\begin{figure}[t]
\includegraphics[width=3in]{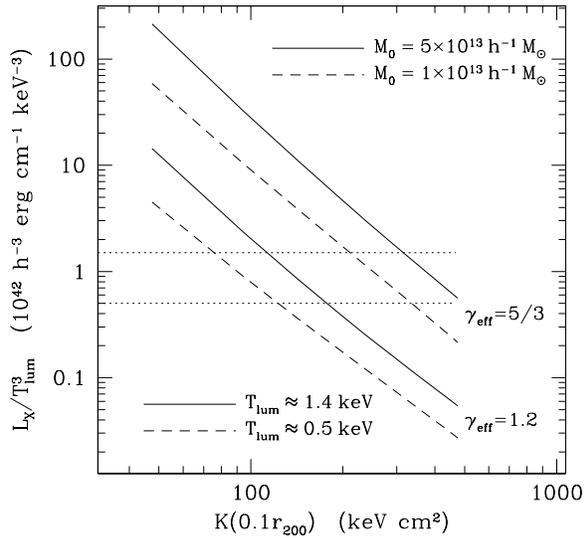}
\caption{ \footnotesize
Dependence of $L_X/T_{\rm lum}^3$ on the entropy
scale $K_{0.1}$ at $0.1 r_{200}$ in polytropic models.  
Tracks are shown for two different halo masses: 
$5 \times 10^{13} \, h^{-1} \, M_\odot$ (solid lines) 
and $1 \times 10^{13} \, h^{-1} \, M_\odot$ (dashed lines).  
Each pair of tracks corresponds to different effective polytropic
index.  The upper tracks represent isentropic models ($\gamma_{\rm eff} 
= 5/3$), and the lower tracks represent polytropic models with an
entropy gradient similar to those observed in clusters ($\gamma_{\rm eff} 
= 1.2$).  The halo concentration is $c_{200} = 10$ in all 
models, and the pressure at $r_{200}$ is the same in all models.  
Dotted lines indicate the same values of $L/T^3$ as in Figure~\ref{lt3rel}. 
One can reproduce the observed $L/T^3$ ratio with
either $\gamma_{\rm eff} \approx 1.2$ and $K_{0.1}
\approx 130 \, {\rm keV \, cm^2}$ or with $\gamma_{\rm eff} 
\approx 5/3$ and $K_{0.1} \approx 350 \, 
{\rm keV \, cm^2}$.
\label{polyfloor}}
\end{figure}

Figure~\ref{lt3rel} shows the luminosity-temperature relation
we need to reproduce.  We plot the quantity $L_X / 
T_{\rm lum}^3$, where $T_{\rm lum}$ is the luminosity-weighted
temperature, because it is nearly constant at
$\sim 1.5 \times 10^{42} \, h^{-3} \, {\rm erg \, s^{-1} \, keV^{-3}}$
for $T_{\rm lum} \gtrsim 1\, \keV$ and a little bit smaller below that
temperature.  Figure~\ref{polyfloor} shows the same
quantity drawn from the polytropic models for two different
halo masses, $M_{200} = 5 \times 10^{13} \, h^{-1} \, M_\odot$
and $M_{200} = 1 \times 10^{13} \, h^{-1} \, M_\odot$ with
luminosity-weighted temperatures of $T_{\rm lum} \approx 
1.4 \, \keV$ and $T_{\rm lum} \approx 0.5 \, \keV$, respectively.
The halo concentration is $c_{200} = 10$  
in all cases and we truncate the
integrations for $L_X$ and $T_{\rm lum}$ at the virial
radius.

% ----------------------------------------
% ----------------------------------------

% ----------------------------------------
% ----------------------------------------

Comparing Figure~\ref{lt3rel} with Figure~\ref{polyfloor} shows
that both the isentropic models and the $\gamma_{\rm eff} = 1.2$ models 
can reproduce the $L$-$T$ relation, but the isentropic models call 
for higher levels of core entropy.
The reason for this behavior is that the density profile 
of the isentropic models is shallower than $\rho \propto 
r^{-3/2}$, so that $L_X$ is dominated by 
emission from near the virial radius.
The overall luminosity is then set by the entropy level
in the outer regions of the group.  Conversely, when
$\gamma_{\rm eff} \approx 1.2$, the density profile
is steeper than $\rho \propto r^{-3/2}$.  The inner
regions then dominate the total luminosity, and the
core entropy can be lower because a smaller amount of
gas produces most of the emission.  Because observations of 
core entropy in $\sim$1~keV halos indicate that $K_{0.1}
\approx 100$-$150 \, {\rm keV \, cm^2}$ (Ponman \etal 1999), 
models with significant entropy gradients, consistent with 
$\gamma_{\rm eff} \sim 1.2$, are probably
closer to the actual entropy profiles of groups.

\subsection{Entropy at the Outskirts}
\label{sec-outskirts}

The preceding analysis suggests that groups
are not isentropic but instead have entropy
gradients more like those observed in clusters.
This finding, when coupled with the elevated 
levels of core entropy observed in groups 
(e.g. Ponman \etal 1999), implies that
entropy at the outskirts of groups must
also be elevated above the predictions of
pure gravitational structure formation.
If the shape of the entropy gradient is 
unchanged, then the main difference between
groups and clusters must be in the normalization
of the entropy gradient rather than its slope.

\begin{figure}[t]
\includegraphics[width=3in]{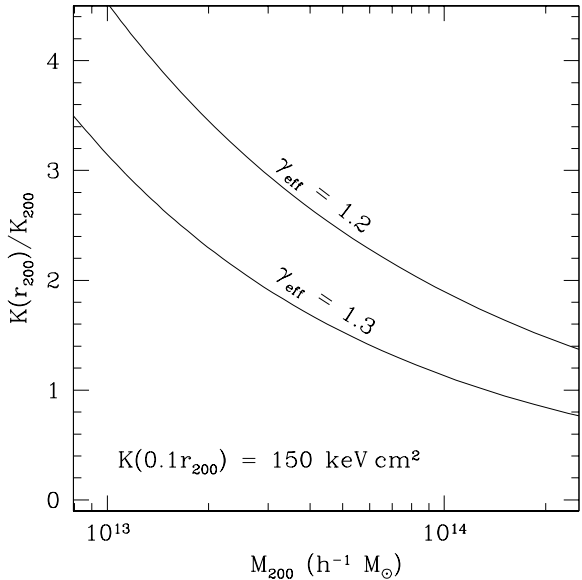}
\caption{ \footnotesize
Entropy at $r_{200}$ in polytropic models with core entropy
$K_{0.1} = 150 \, \keV \, {\rm cm^2}$.  
The polytropic indices ($\gamma_{\rm eff} \sim 1.2 - 1.3$)
consistent with both the $L$-$T$ relation and the observed core entropy 
of groups $\sim 100$-$150 \, {\rm keV cm^2}$, imply $K(r_{200})
\sim 2$-$3 K_{200}$.  
\label{polymax}}
\end{figure}

Figure~\ref{polymax} shows the scaled entropy 
$K/K_{200}$ at $r_{200}$ implied by polytropic 
models having different halo masses but a fixed
value of core entropy $K_{0.1} = 150 \, \keV \, 
{\rm cm^2}$.  In polytropic models with $\gamma_{\rm eff} = 1.2$
and this value of core entropy, the implied entropy
at $r_{200}$ is $\sim 2$-4 times larger than $K_{200}$ 
over the mass range $10^{13} \, h^{-1} \, M_\odot$ 
to $10^{14} \, h^{-1} \, M_\odot$.  For comparison,
we also show the implied entropy for an effective polytropic 
index of $\gamma_{\rm eff} = 1.3$, which agrees less well
with the observational constraints.  In that
case, the outer entropy levels drop to $\sim 1$-3 times $K_{200}$.
However, the entropy levels at $K_{200}$ in groups simulated
without preheating or cooling are rarely much greater
than $K_{200}$, with no significant
trend in mass (see Figure~\ref{L50_nf}).  Notice
that these entropy enhancements are quite difficult
to produce by heat input alone. The value of $K_{200}$
for a $10^{14} \, h^{-1} \,M_\odot$ halo is $\sim 700 \, \keV
\, {\rm cm}^2$, so doubling that entropy would require an
enormous amount of heat.

% ----------------------------------------
% ----------------------------------------

Direct observations of entropy at the outskirts of a few
groups support the idea that groups have significant entropy
gradients and higher maximum entropy levels than predicted
by self-similar models that do not include non-gravitational
processes.  Finoguenov \etal (2002) determined
$Tn_e^{-2/3}$ as a function of radius in several groups 
using {\em ASCA} and {\em ROSAT} data and compared them
with the entropy gradients of more massive systems. 
After dividing these gradients by $M_{500}^{2/3}$ to
remove the dependence on mass scale, they found that
the outer entropy levels of groups exceeded those in
clusters by a similar factor of 2-3, with a large scatter.

Finoguenov \etal (2002) interpreted this excess entropy
as a $\sim 400 \, {\rm keV \, cm^2}$ ``entropy ceiling''
produced entirely by non-gravitational heating owing to
galactic winds.  In this interpretation, the 
entropy gradient results from a gradual rise in
the entropy of the intergalactic medium external
to a group before it accretes onto the group.
For this level of entropy to be produced by
supernovae, the energy injection must occur with
near 100\% efficiency at redshift $z \sim 3$ into
regions of relatively low overdensity ($\sim 5 
\rho_{\rm cr}$).

\subsection{A Smooth-Accretion Interpretation}
\label{sec-smoothinterp}

We would like to offer another interpretation of this
apparent entropy excess that is based on the smooth accretion 
models developed in this paper.  Relatively small
amounts of preheating ($\lesssim 100 \, {\rm keV \,
cm^2}$) can eject much of the gas from the smaller
halos that accrete onto a group or a large elliptical
galaxy at late times, making the accreting gas more
homogeneous.
%\footnote{After submitting this paper, we
%received a preprint by Ponman \etal proposing a similar 
%interpretation for the entropy excess in which preheating 
%lowers the density of gas accreting through filaments.}
This more modest level of preheating can
therefore break self-similarity by allowing the group 
to generate most of its entropy through smooth accretion.  
The general shape of the entropy profile remains similar
to the lumpy-accretion case because it is determined by
the accretion history (\S~\ref{sec-acchist}), but its normalization
is boosted because the mean density of incoming gas is
reduced (\S~\ref{sec-depsmooth}).

In the limit of perfectly smooth accretion, 
equation~(\ref{eq-khatcold}) implies that
\begin{eqnarray}
  \frac {K(r_{200})} {K_{200}} \approx  
        2.6 \, \left( \frac {d \ln M} {d \ln t} \right)^{-2/3} \; \; ,
\label{eq-entenhance}
\end{eqnarray}
assuming that $H_0 t_0$, $2 \xi$, and $\eta$ are all $\approx 1$.
The enhancement of entropy at $r_{200}$ relative to expectations
from hierarchical accretion is larger for smaller halos because
it depends on the mass accretion rate.  Inserting the $d \ln M /
d \ln t$ values at $t=t_0$ from Figure~\ref{mtfits} into equation
(\ref{eq-entenhance}) yields $K(r_{200})/K_{200} \approx 3.5$
for $10^{13} \, h^{-1} \, M_\odot$, $\approx 2.4$
for $10^{14} \, h^{-1} \, M_\odot$, and $\approx 1.7$
for $10^{15} \, h^{-1} \, M_\odot$.
Note also that smooth accretion can increase the 
effectiveness of preheating at limiting condensation in groups, 
as the initial entropy input needed to smooth the accreting gas 
is strongly amplified at the accretion shock.

\section{The Puzzle of Entropy Normalization}
\label{sec-puzzle}

Groups and clusters appear to have similar entropy profiles,
consistent with an effective polytropic index $\gamma_{\rm eff}
\sim 1.2$ at large radii.  However, the normalizations of those
profiles cannot scale with $K_{200} \propto M_{200}^{2/3}$,
as they would if they were determined by purely gravitational
processes.  Instead, the relationship between entropy 
normalization and halo mass is somewhat shallower, quite
possibly because preheating has partially smoothed the
intergalactic medium.  Thus, we would like to understand
more quantitatively how entropy production depends on
the homogeneity of accreting gas.

The normalization of intracluster entropy in the
smooth-accretion limit is relatively simple to
calculate (see \S~\ref{sec-smoothcold}), but the
physics that sets the normalization during lumpy
accretion is more complex.  This section outlines
some of the physical processes that might govern
entropy production by lumpy accretion.  We first
present a naive model for merger shocks in which 
all the accreting gas belongs to virialized subhalos 
and all the incoming kinetic energy is thermalized 
through merger shocks within 
those subhalos.  This naive model fails to produce
the required amount of entropy, implying that 
either (1) the preshock density distribution
and perhaps the accretion velocity assumed in 
the naive model poorly represent reality, or 
(2) some of the incoming kinetic energy is thermalized
within the intracluster medium of the main halo.
One way to account for the first possibility is 
to adjust the preshock density distribution,
and we present an illustrative example involving
power-law density distributions.  The second
possibility implies that lumpy accretion provides a
distributed source of heating throughout a cluster,
and we conclude the section by identifying the
hallmarks of distributed heating and suggesting
how to quantify it in numerical simulations.

\subsection{Naive Merger Shocks}
\label{sec-discrete}

The accretion process usually envisioned in semi-analytic models 
of cluster formation calls for all the accreting gas to lie within 
a subhalo of some kind, even if many of those subhalos sit below 
the practical resolution limit of the calculation (e.g., Wu, 
Fabian, \& Nulsen 2000; Bower \etal 2001).  
In a self-consistent semi-analytical model one would like to
track how entropy develops as these subhalos collide and merge
with the main halo and to calculate the density configuration of
the new halo from that evolving entropy distribution 
(e.g., Balogh \etal 2003, in preparation).  Instead, the 
density distribution is typically assumed to settle into a 
polytropic or $\beta$-model configuration similar to 
observed density profiles, with a core temperature equal 
to the virial temperature.  Here we show that 
accretion shocks within discrete, accreting subhalos 
do not generate enough entropy to reproduce the 
characteristic entropy profile of a cluster.

If the mean gas density within each of the accreting subhalos
is $\Delta f_b \rho_{\rm cr}$ and strong shocks thermalize 
all the kinetic energy of accretion within those subhalos
as they cross the virial radius of the main halo, then
for $\xi \approx 0.5$ we find that the mean postshock
entropy of gas accreting at time $t$ is
\begin{eqnarray}
\label{eq-haloacc}
  \bar{K}(t)  & \approx & \left( \frac {\rho_1} 
             {\Delta f_b \rho_{\rm cr}} \right)^{2/3}  K_{\rm sm}(t)\\
      ~ & \approx & \left[ \frac {1} {3}
                      \left( \frac {2} {\Delta} \right)^{1/2}
                      (Ht)^{-1}
                      \frac {d \ln M} {d \ln t} 
                      \right]^{2/3}   K_{\rm sm}(t) \nonumber \; \; .
\end{eqnarray}
The cosmologies and accretion histories considered here give
$\bar{K}(t) \sim 0.1 (\Delta/200)^{-1/3} K_{\rm sm}(t)$, 
several times smaller than the value found in
the simulations and observations.  According to \S~\ref{sec-entjump},
about 84\% of the preshock entropy in the subhalos should be
added to $\bar{K}$, but this does not come close to fixing 
the problem unless the accreting halos are similar in size 
to the main halo, which is rare during the formation of a
large cluster.  

This postshock entropy deficit can be recast in terms of
the shock velocity needed to rectify it.  Suppose that
the accreting subhalos have gas density distributions similar 
to that of the main halo.  Because the gas temperature
in the main halo is $\approx GM \mu m_p /2r_\Delta$, boosting the
entropy distribution of the subhalo's gas to match that
of the main halo requires a shock velocity
\begin{eqnarray}
 v_{\rm sh} & \approx & \left[ 
                          \frac {3 \cdot 4^{2/3}} {2}
                          \frac {G M} {r_\Delta} 
                          \right]^{1/2} 
                          \nonumber \\
       ~   & \approx & 2 v_{\rm ac} \; \; ,
\end{eqnarray}
where $v_{\rm ac}$ is the infall velocity
at radius $r_\Delta = r_{\rm ta} / 2$.
A gas blob entering a cluster with an NFW potential would 
need to fall unimpeded from its turnaround radius to 
$\lesssim 0.25 r_\Delta$ before it reached a velocity
$\approx 2 v_{\rm ac}$.  However, in the naive model
we have outlined, infalling blobs do not fall so far 
toward the center of the main halo before being
shocked and disrupted.  If a dense lump of accreting
gas does fall this far, then dissipation of its kinetic
energy through shocks and turbulence is also quite likely 
to heat the main halo's gas.  We consider how this 
mechanism affects entropy generation in \S~\ref{sec-heating}.

\subsection{Adjusted Merger Shocks}
\label{sec-vfilling}

If one retains the assumption that merger shocks thermalize
all the incoming energy within the accreting gas, then
then one must adjust the incoming density distribution
and perhaps raise the shock velocity.  Striking the
proper blend of smooth accretion and lumpy accretion
would then yield the correct entropy level.  
To illustrate this idea, we briefly outline a case
in which the accreting gas fills the accreting volume
instead of being confined to regions with mean density
$\sim \Delta f_b \rho_{\rm cr}$.

Suppose that the preshock density of accreting gas is 
distributed like a power law.  Define $f(\rho) d \rho$ 
to be the fraction of accreting gas with 
density between $\rho$ and $\rho+d\rho$, so that 
$f(\rho) = (p-1) \rho_{\rm min}^{p-1} \rho^{-p}$ for a
power-law index $p > 1$.  If we then assume
that all the gas accreting at a given time
shocks at the same accretion radius, we can integrate 
$\rho^{-2/3} f(\rho) d\rho$ to find the mean entropy 
of gas accreted at time $t$: 
\begin{equation}
  \bar{K}(t) = \frac {p-1} {p- \frac {1} {3}} \left(
                      \frac {\rho_1} {\rho_{\rm min}} \right)^{2/3}
                       K_{\rm sm}(t)  \; \; .
\end{equation} 
If we further assume that the accreting gas fills the available 
volume, then $\rho_1 / \rho_{\rm min} = p / (p-1)$ and
\begin{equation}
  \bar{K}(t) = \frac {p^{2/3} (p-1)^{1/3}} {p- \frac {1} {3}}  K_{\rm sm}(t)
        \; \; .
\end{equation}

The power-law index $p$ is directly related to the density distribution
within an accreting subhalo.  In particular, for $\rho(r) \propto 
r^{-q}$, we have $p = 3/q$.  The case of extended singular 
isothermal spheres accreting onto the main halo therefore 
corresponds to $p = 3/2$, and we get $\bar{K} \approx
0.89 K_{\rm sm}(t)$.  In the limit of small $q$ and large $p$, the
incoming density distribution is nearly uniform, and $\bar{K} \approx
K_{\rm sm}(t)$.  However, $\bar{K}$ grows
arbitrarily small in the limit of $q \rightarrow 3$ and
$p \rightarrow 1$, in which case the majority of the gas 
becomes concentrated at high density.  Thus, for some
value of $q \lesssim 3$, the mean postshock entropy has 
the desired value.

This calculation implies that one can specify a preshock density
distribution that yields the correct postshock entropy distribution 
after passing through an accretion shock.  However, this model
for entropy production is not necessarily the solution to the entropy
normalization puzzle.  In order to test it, one would need to
measure the density distribution and infall velocity of 
baryonic material accreting onto a numerically simulated 
cluster and then compute the postshock entropy distribution
implied by this model.

\subsection{Distributed Heating}
\label{sec-heating}

An alternative solution to the entropy normalization problem
depends on an important qualitative difference between 
smooth accretion and lumpy accretion.
Smooth accretion thermalizes all the incoming kinetic energy
at the accretion shock, within gas that has just accreted.  
After a gas shell has become part of the intracluster medium, 
its entropy is no longer affected by the smooth-accretion process.  
However, dense lumps of accreting gas can penetrate the 
virial radius without substantially decelerating.  They 
subsequently move through the main cluster, stirring up
turbulence and stimulating shocks within the
main cluster's gas.  These processes continue until either
drag saps all of the dense lump's kinetic energy or
hydrodynamic instabilities tear the dense lump apart.

The recently observed presence of cold fronts in clusters
(Markevitch \etal 2000; Vikhlinin, Markevitch, \& Murray
2001) suggests that dense subhalos can remain coherent
for quite a long time before blending with the rest of
the intracluster medium.  Thus, much of the incoming 
kinetic energy of lumpy accretion may ultimately be
deposited into the pre-existing intracluster medium
as the disturbances stirred up by the accreting lumps 
dissipate and thermalize (see e.g., Gomez \etal 2002).  
If that is the case, then accretion provides a distributed
source of heating that adds entropy to the entire intracluster 
medium, not just to gas that has recently accreted.  

Here we examine  some crucial differences
between entropy production through smooth accretion 
and entropy production through distributed heating.  
First we assess the amount of entropy 
growth that lumpy accretion needs to generate
in order to maintain the self-similar entropy profiles
observed in simulations and in hot clusters.  Then
we examine the implications of heating the existing 
intracluster medium through lumpy accretion, showing 
that the source of condensing low-entropy gas in the 
lumpy-accretion mode is different from its origin
in the smooth-accretion mode.
If distributed heating is important, then much
of the core gas in present-day clusters may come
from the cores of subhalos that accrete at $\sim 0.5
t_0$.  One can assess the importance of distributed
heating by tracking the $K(t)$ trajectories of 
Lagrangian gas parcels in numerical simulations 
after they accrete onto a cluster.

\subsubsection{Self-Similar Entropy Growth}

One notable feature of numerically modelled 
clusters in simulated CDM-like cosmologies is 
their near self-similarity.  The dark-matter potential
wells approximately follow the NFW density profile, and
the main systematic devations from self-similarity can
be characterized with a halo concentration parameter that
increases with decreasing halo mass (e.g., Navarro \etal 1997).  
This dependence on halo mass comes about because smaller 
halos tend to accumulate their mass earlier in time, 
when the mean density of the universe is larger (Navarro \etal 
1997; Bullock \etal 2001; Eke, Navarro, \& Steinmetz 2001).
Numerical models including hydrodynamics also show that
the baryonic density and temperature profiles in the absence 
of non-gravitational heating and cooling processes 
closely track the analogous dark-matter profiles, except 
in the very center (e.g., Navarro, Frenk, \& White 1995;
Eke \etal 1998; Frenk \etal 1999).  
Therefore, whatever processes are responsible for 
entropy growth during hierarchical structure formation 
act to preserve this near self-similarity.

If clusters were precisely self-similar, then their properties
would be entirely determined by their mass $M$ and the time
of observation, which determines the mean mass density $\Delta 
\rho_{\rm cr}$ within the cluster.  According to the simple spherical 
infall model of \S~\ref{sec-accrad}, the turnaround radius of matter
accreting at time $t$ is $r_{\rm ta} = (2GMt^2/\pi^2)^{1/3}$
and it falls through the virial radius ($r_{\rm ta} / 2$) with
a kinetic energy per particle $T_\phi = GM \mu m_p/ r_{\rm ta}$.
(These expressions are strictly true only if $\Lambda = 0$, but
they are accurate to within 10\% for $\Lambda$CDM with 
$\Omega_\Lambda = 0.7$.)  Because the mean density within 
the virial radius is $3 \pi / G t^2$,
the entropy scale\footnote{This entropy scale exceeds $K_{200}$
by a factor $\approx (200/\Delta)^{2/3}$ or $\approx 1.6$ for
a $\Lambda$CDM cosmology at the present time.  We are using
$K_\phi$ in the self-similar analysis because it depends
explicitly on $M$ and $t$ alone.} associated with a 
cluster's baryons is 
\begin{equation}
  K_\phi = \frac {1} {2} \left( \frac {2G^2} {3 f_b} \right)^{2/3}
                     (Mt)^{2/3} \; \; .
\end{equation}
The dimensionless entropy distribution function $K(M_g)/K_\phi$
would then be the same for all clusters.  

One can integrate over this distribution to find the total
classical thermodynamic entropy of a cluster's baryons,
\begin{equation}
  S(M,t) = \frac {f_b M} {\mu m_p} \left[ \ln (Mt) 
            + {\rm const.} \right] \; \; ,
\end{equation}
where the details of the self-similar distribution 
function are absorbed within the constant term.  
If the main halo accretes a subhalo of mass 
$\delta M \ll M$ during the time interval $\delta t \ll t$,  
then the entropy of the entire system, including 
the pre-accretion entropy of the subhalo's baryons, 
rises by 
\begin{eqnarray}
 \delta S & = & S(M+\delta M,t+\delta t) - S(M,t) \nonumber \\
     ~   & ~ &  \; \; \; \; \; \; \; \; \; - \; S(\delta M,t) 
                  \nonumber \\
     ~    & = & \frac {f_b \delta M} {\mu m_p} 
                 \left[ 1 + 
                 \frac {M} {t} \frac {\delta t} {\delta M} 
                   + \ln \left( \frac {M} {\delta M} \right) \right] \; \; .
\end{eqnarray}
The first two terms in the lower expression correspond to
the heat input needed to maintain self-similarity in
the merged system, which presumably comes from thermalization
of incoming kinetic energy.  The third term corresponds
to the entropy associated with raising the subhalo's gas
to the temperature of the main halo.  The most efficient 
way to increase the entropy of that gas is to raise its 
temperature while keeping its density constant, perhaps 
through conduction of heat from the main cluster facilitated by
rapid mixing.\footnote{Heat input that 
raises the temperature of a monatomic ideal gas from $T_1$ 
to $T_2$ at constant density increases its specific entropy 
$s$ by $(3/2) \ln (T_2/T_1)$. The temperature difference
between self-similar halos is $T_2/T_1 = (M/\delta M)^{2/3}$,
corresponding to a specific entropy increase $\ln (M/\delta M)$.}  
However, the subhalo's baryons are more likely to be 
compressed or shock heated before mixing with the main 
intracluster medium, in which case a portion of the third 
term corresponds to thermalization of incoming 
kinetic energy within the subhalo's gas.

Smooth accretion achieves self-similarity in a different way.
From equation~(\ref{eq-khatcold}) it is evident that cold accretion
can produce self-similar clusters as long as $\zeta \equiv
d \ln M / d \ln t$ is the same for all clusters; the entropy
profile produced by smooth accretion for constant $\zeta$ 
is $K(M_g,t) = K_{\rm sm}(t) [M_g/f_b M(t)]^{2/3+2/3\zeta}$.
In that case, the classical thermodynamic entropy of 
a cluster's baryons is
\begin{equation}
  S = \frac {f_b M} {\mu m_p} 
              \left[ \ln K_{\rm sm}^{3/2} - 
                  \left( 1 + \frac {1} {\zeta} \right) + s_0 \right]
                  \; \; ,
\end{equation}
which explicitly includes the constant term arising from the 
cluster's internal structure.  Smooth accretion of gas mass 
$f_b \delta M$ with pre-accretion entropy $K_1 \ll K_{\rm sm}$ 
during a time interval $\delta t$ therefore adds
\begin{equation}
 \delta S = \frac {f_b \delta M} {\mu m_p} 
              \left[ \frac {3} {2}
                      \ln \left( \frac {K_{\rm sm}} {K_1} \right)
                             \right]
\end{equation}
to the thermodynamic entropy of the whole accreting system,
after the initial entropy of the accreting gas has been subtracted.
The entropy increase in this case is due entirely to shock
heating of incoming gas at the accretion shock, and there
are no additional heating terms.

\subsubsection{Implications of Accretion Heating}

The preceding analysis suggests that hierarchical merging
produces entropy exceeding the amount needed to raise the
accreting gas to the temperature of the original halo.
A detailed accounting of how this heating occurs is
beyond the scope of this paper, but we would like to
suggest a qualitative picture that can be tested
and refined with numerical simulations.  Here we 
construct a toy model for distributed heating 
owing to accretion and show how it affects the evolution of
the intracluster entropy distribution.  If distributed
heating is the dominant mode of entropy growth,
then the entropy of gas deep within a cluster
should rise with time, possibly compensating
for some of the radiative cooling of the
cluster core.

Let the rate at which accretion deposits heat energy
into the intracluster medium be $\epsilon 
T_\phi f_b \dot{M} / \mu m_p$.  If this energy deposition
rate were equal to the rate at which kinetic energy
passes through the virial radius, then we would have
$\epsilon = 1$.  However, accreting gas that is denser 
than gas at the virial radius plunges to smaller radii, 
gaining additional energy to be thermalized as it 
descends into the cluster.  A dense gas blob eventually 
comes to rest when its entropy equals that of the 
surrounding gas.  Thus, the total rate of heat input 
owing to lumpy accretion might be somewhat larger than 
the rate at which kinetic energy flows through the 
virial radius, allowing $\epsilon$ to exceed unity.

Distributing this heat input equally among all the
cluster's gas particles leads to the following expression 
for the evolution of entropy with time within a Lagrangian 
gas parcel, including radiative cooling:
\begin{equation}
 \frac {d \ln K^{3/2}} {dt} =  
      \frac {\epsilon T_\phi} {T} \frac {d \ln M} {d \ln t} \frac {1} {t} 
      - \frac {3} {2} \frac {K^{3/2}_c(T)} {K^{3/2}} \frac {1} {t_0}
        \; .
  \label{eq-entevol}
\end{equation}
A more general model could be constructed by allowing
$\epsilon$ to depend on $K$,
thereby accounting for inhomogeneities in heat input.
The mean value of $\epsilon$, integrated over time,
and its spatial variations within a cluster would be
interesting to measure in numerical simulations of
hierarchical merging.

% ----------------------------------------
\begin{figure}[t]
\includegraphics[width=3in]{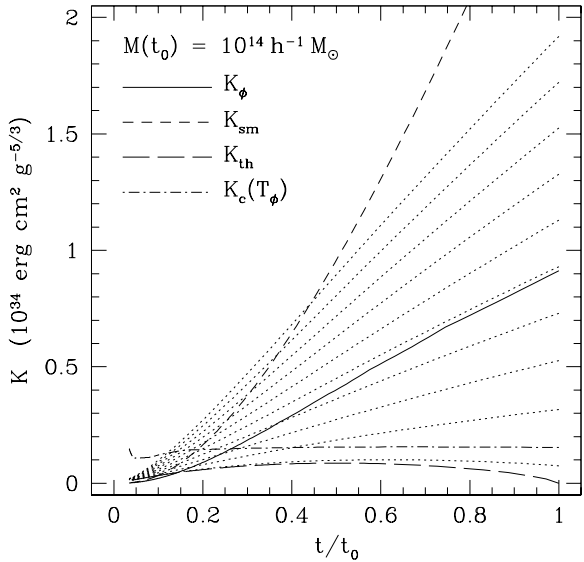}
\caption{ \footnotesize
Change in entropy with time in the intracluster medium of the
main halo owing to distributed heating and radiative cooling.
The dotted lines show how Lagrangian gas parcels move through the
$K$-$t$ plane, according to equation~\ref{eq-entevol} with 
$\epsilon = 1 + (d \ln M / d \ln t)^{-1}$
and $T = T_\phi$.  The solid line indicates the characteristic
entropy $K_\phi(t)$ of the growing dark-matter halo.  The short-dashed 
line gives $K_{\rm sm}(t)$ for $\xi = 0.5$ and no preheating.  
The long-dashed line shows the threshold $K_{\rm th}$ below which
a gas parcel is destined to condense by $t = t_0$.  The dot-dashed
line shows the cooling threshold $K_c(T_\phi)$ corresponding to the
characteristic temperature $T_\phi(t)$ of the halo.  Because of
the heat input generated by accretion, the entropy of the intracluster
medium can increase with time, amplifying the effects of early
entropy input and reducing the amount of gas that condenses by 
$t = t_0$.
\label{ent_plane_hc14}}
\end{figure}
% ----------------------------------------

% ----------------------------------------
\begin{figure}[t]
\includegraphics[width=3in]{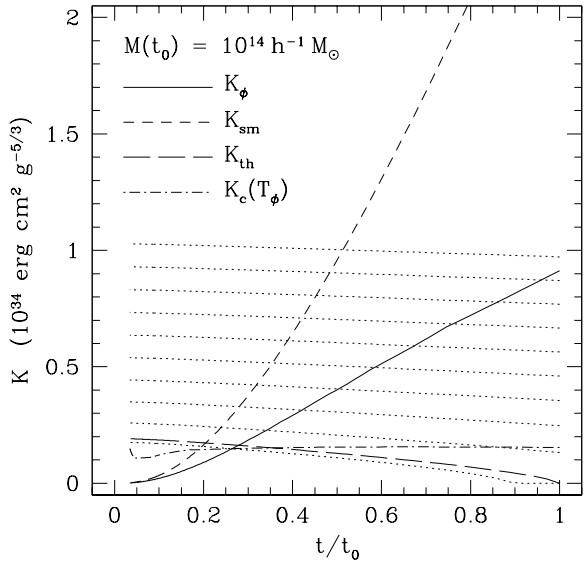}
\caption{ \footnotesize
Change in entropy with time in the intracluster medium of the
main halo owing to radiative cooling only.  The lines have the
same meanings as in Figure~\ref{ent_plane_hc14}.  In this scenario,
gas that accretes early in time is much more likely to condense
by $t = t_0$.
\label{ent_plane_c14}}
\end{figure}
% ----------------------------------------

For purposes of illustration, let us consider an extreme case
of distributed heating explicitly tuned to preserve self-similarity.
If $T \approx T_\phi$ and $K_c \ll K$, the cluster will remain 
self-similar as long as
\begin{equation}
  \epsilon \approx 1 + \left( \frac {d \ln M} {d \ln t} \right)^{-1}
       \; \; .
\label{eq-epsilon}
\end{equation}
Figure~\ref{ent_plane_hc14} shows how the value of $K$ associated with
a Lagrangian gas parcel changes with time in a cluster of mass
$10^{14} \, h^{-1} \, M_\odot$ at $t_0$ according to equation
(\ref{eq-entevol}) with $\epsilon$ set by equation (\ref{eq-epsilon}), 
assuming the accretion history from Tozzi \& Norman (2001).  For simplicity,
we have assumed $T = T_\phi$ throughout the cluster.  The
dotted lines indicate how $K(t)$ of a gas parcel evolves
after it appears in a certain position in the $K$-$t$ plane.
Gas parecels that have become incorporated into the cluster
should evolve along these paths; however, the starting points
of the paths are arbitrary because this model does not specify
the entropy of a gas parcel just after it has accreted.
The solid line shows $K_\phi(t)$, which runs approximately 
parallel to the neighboring $K(t)$ tracks because we have chosen
$\epsilon$ to reproduce self-similar entropy growth.

In this scenario, distributed heating owing to accretion
is quite important early in the cluster's history.  The 
long-dashed line labeled $K_{\rm th}$ corresponds to the
$K(t)$ track that ends at $K(t_0) = 0$.  Gas blobs below 
this line are destined to cool and condense by the present 
time, unless internal feedback intervenes.  Notice that
a relatively small amount of initial entropy ($< 10^{33}
\, {\rm erg \, cm^2 \, g^{-5/3}}$) can prevent intracluster
gas from condensing by the present day, because distributed
heating acts as an entropy amplifier.

The situation is quite different in Figure~\ref{ent_plane_c14}, 
in which we have computed $K(t)$ with $\epsilon = 0$.  All 
the gas with initial entropy $\lesssim 2 \times 10^{33} \, {\rm erg 
\, cm^2 \, g^{-5/3}}$ at early times condenses by the present time.  
If smooth, cold accretion is the dominant mode, then gas 
parcels enter the $K$-$t$ plane on the short-dashed line 
labeled $K_{\rm sm}$, and everything that accretes before
$0.2 t_0$ will radiate away all of its entropy.  Substantial
amounts of preheating or internal feedback ($\gtrsim 2 \times 
10^{33} \, {\rm erg \, cm^2 \, g^{-5/3}}$) are necessary to 
prevent this gas from condensing.

In both scenarios, the fate of a gas parcel accreting onto
the cluster depends on where it enters the $K$-$t$ plane
as it becomes incorporated into the intracluster medium.
For example, if the core of an accreting subhalo does
not gain enough entropy during the merger to exceed
$K_{\rm th}(t)$, then it will ultimately sink toward
the center of the cluster and condense, unless some
other heat source prevents it from doing so.  The level of this 
threshold depends critically on $\epsilon$, which
is why this parameter should be measured using numerical 
simulations.  Sampling of the core gas in the
simulations of Navarro \etal (1995) shows that $K(t)$
indeed rises with time (see their Figure~8), implying 
$\epsilon > 0$, but the data presented are insufficient 
for measuring the heating rate.

%If $\epsilon$ is close to the value
%inferred from self-similar entropy growth, then the 
If $\epsilon$ is close to the similarity-preserving
value assumed in equation (\ref{eq-epsilon}), then the
origin of ``cooling flow'' gas in hierarchical merging
differs substantially from the smooth accretion case, 
in which the lowest-entropy gas is the earliest gas to 
accrete.  Large values of $\epsilon$ strongly boost the 
entropy of gas accreted at early times, preventing 
it from cooling later on.  Instead, $K_{\rm th}(t)$
reaches its maximum value near $t_0/2$, suggesting that
the gas most likely to be part of a cooling flow is 
the gas that accretes with low entropy at intermediate 
times.

Unfortunately, the simple model we have developed provides 
little insight into where a parcel of accreting gas should
enter the $K$-$t$ plane after a merger.  Shocks 
probably raise the entropy of this gas somewhat before 
it mixes with the intracluster medium, but the magnitude
of this entropy increase is uncertain.  Tracking the
entropy of gas parcels during a numerically simulated
merger would help to quantify this important parameter.

\section{Summary}
\label{sec-summary}

The observable X-ray properties of a relaxed cluster of
galaxies are entirely determined by the entropy distribution
of its intracluster medium and the shape of the potential
well that confines that medium.  Because intracluster entropy
is so fundamental, we have sought to understand its origin
in terms of some simple analytical models.  This analysis
leads us to conclude the following:
\begin{enumerate}

\item A classical smooth accretion shock produces 
      an intracluster entropy distribution with a
      form much like the self-similar entropy
      profiles observed in both massive clusters 
      and in simulations of hierarchical structure
      formation.  However, the normalization of the
      entropy profile arising from smooth accretion is
      $\sim$2-3 times higher, depending on the accretion 
      rate.  Smooth, steady, isotropic accretion produces the
      maximum amount of entropy for a given $\dot{M}$ 
      because smoothing of the incoming gas minimizes
      the mean mass-weighted density entering the
      accretion shock (\S~\ref{sec-smoothcold}). 
       
\item Preheating smoothly accreting gas to an entropy
      level $K_1$ boosts the postshock entropy
      profile by an additive term $\approx 0.84 K_1$,
      as long as the entropy of preheating does not
      exceed the characteristic entropy of the 
      accreting halo (\S~\ref{sec-entjump}).
      However, preheating that exceeds the characteristic
      entropy of the accreting halo strongly suppresses
      entropy production at the accretion shock
      because it inflates the intracluster medium,
      pushing the accretion shock well outside the
      virial radius and thereby reducing the shock
      velocity at the accretion front (\S~\ref{sec-preheat}).
      
\item Because smooth accretion of cold or moderately
      preheated gas produces $\sim$2-3 times more
      entropy than observed in both massive clusters 
      and simulations, inhomogeneity of the incoming 
      gas must play a role in setting the entropy level 
      of the intracluster medium. 
      This difference in entropy production between smooth 
      accretion and hierarchical accretion allows for an interesting
      mode of similarity breaking. If preheating
      has been able to eject gas from the small subhalos
      accreting onto groups, then accretion of baryons
      onto groups may be smoother than accretion onto
      clusters, enhancing entropy production at the
      accretion shock (\S~\ref{sec-inhomopre}).
    
\item Smoothing of gas associated with low-mass halos might 
      explain some interesting differences between the
      entropy profiles of groups and clusters.
      Isentropic groups cannot satisfy  
      the observed $L$-$T$ relation while 
      having the observed core entropy $\sim 100$-$150 \,
      {\rm keV \, cm^2}$ (\S~\ref{sec-polyfloor}).
      Polytropic models of groups that satisfy both 
      observational constraints have entropy gradients
      implying entropy levels at $\sim r_{200}$ that substantially
      exceed what hierarchical accretion can produce
      (\S~\ref{sec-outskirts}).
      Direct observations indicate that entropy at
      the outskirts of groups is indeed $\sim$2-3 times 
      higher than values derived from self-similar scaling 
      of clusters (Finoguenov
      \etal 2002), perhaps owing to a transition
      from lumpy accretion to smooth accretion
      occuring on group scales (\S~\ref{sec-smoothinterp}).

\item Because the normalization of intracluster entropy
      appears to be sensitive to the density distribution
      of incoming gas, we would like to understand what sets
      that normalization in the case of hierarchical accretion.
      However, a naive model in which all the incoming energy
      is thermalized by shocks within subhalos having an 
      average density similar to that of the main halo
      fails to produce enough entropy (\S~\ref{sec-discrete}).
   
\item Given the failure of this simple model to explain
      the observed normalization of the intracluster
      entropy distribution, we are driven to consider
      more complex scenarios for entropy generation.
      One possibility is that some of the accreting gas
      is not contained within subhalos, so that the incoming
      density distribution is intermediate between smooth
      accretion and the assumptions of \S~\ref{sec-discrete}.  
      Another possibility is that some of the incoming kinetic 
      energy is thermalized within the existing intracluster 
      medium through shocks and turbulence stimulated
      as accreting gas lumps circulate through the cluster
      (\S~\ref{sec-heating}).
      
\item If the kinetic energy of accretion is indeed
      an important source of distributed heating
      within a cluster, then the development of 
      entropy in the lumpy accretion case is qualitatively
      different from the smooth accretion case in which
      all entropy is generated at the accretion shock
      and is deposited exclusively into the accreting 
      gas.  One can test this possibility by measuring
      the rate of entropy growth deep within the
      accretion radii of simulated clusters (\S~\ref{sec-heating}).

\end{enumerate}

The authors would like to thank Paolo Tozzi for
assistance with his models as well as Gus Evrard and
Trevor Ponman for their comments on the original draft.  
GMV received partial support from NASA through grant NAG5-3257.
MLB is supported by a PPARC fellowship.  
RGB acknowledges the support of the 
Leverhulme foundation.  CGL was supported at 
Durham by the PPARC rolling grant in Cosmology
and Extragalactic Astronomy.

\clearpage
\onecolumn

%\begin{appendix}
\begin{center}
APPENDIX A\\ ~ \\
SELF-CONSISTENT SHOCK RADIUS \\
\end{center}

\renewcommand{\theequation}{A\arabic{equation}}
\setcounter{equation}{0}

The position of the shock radius in an intracluster medium
bounded by accretion pressure depends on both the mass accretion
rate and the internal structure of the cluster.  In this appendix,
we develop a useful approximate solution for the shock radius in
terms of the shock-velocity parameter $\xi$,
based on a polytropic model for structure of the intracluster medium.  
The cluster potential is assumed to be of the NFW type, so that
\begin{equation}
  \phi(r) = - \frac {2 T_\Delta} {\mu m_p} 
              \frac {\ln (1+c_\Delta  r/r_\Delta)} 
                {\ln(1+c_\Delta) - c_\Delta (1+c_\Delta)^{-1}} 
              \frac {r_\Delta} {r} \; \; ,
\end{equation}
where $T_\Delta = G M_\Delta \mu m_p / 2 r_\Delta$ and $M_\Delta$
is the total mass within $r_\Delta = r_{\rm ta}/2$.  
We also assume that the pressure and density of gas 
in the equilibrium configuration are related by a polytropic 
equation of state, with $P(r) \propto [\rho(r)]^{\gamma_{\rm eff}}$.  
Simulations, observations, and the smooth-accretion analysis 
of \S~\ref{sec-preheat} all suggest that $\gamma_{\rm eff} 
\approx 1.2$ is an adequate description of the equilibrium 
state (e.g., Markevitch \etal 1998, 1999; Ettori \& Fabian
1999; Finoguenov \etal 2001; Molendi \& De Grandi 2002); 
however, it is important to realize that this relationship 
is not the actual equation of state of the intracluster gas.  
It is merely a convenient description of the global equilibrium 
state which ultimately depends on the entropy distribution 
$K(M_g)$ and the shape of the potential well.

Given these assumptions, the equation of hydrostatic equilibrium
reduces to
\begin{equation}
  \frac {dT} {dr} = - \mu m_p 
          \left( \frac {\gamma_{\rm eff} - 1} {\gamma_{\rm eff}} \right)
          \frac {d \phi} {dr} \; \; ,
\end{equation}
which yields the solution
\begin{eqnarray}
  T(x) & = & T_\Delta g(x)           \label{eq-tg} \\
  \rho(x) & = & \rho_g [g(x)]^{1/(\gamma_{\rm eff} - 1)} \\
  P(x) & = & \frac {T_\Delta \rho_g} {\mu m_p} 
                   [g(x)]^{\gamma_{\rm eff}/(\gamma_{\rm eff} - 1)} \\
  g(x) & = & g_0(x) + g_1 \\
  g_0(x) & = & \frac {2 (\gamma_{\rm eff} - 1)} {\gamma_{\rm eff}} 
               F(c_\Delta) \frac {\ln (1+c_\Delta x)} {x \ln (1+c)} \\
  F(c) & = & \frac {\ln (1+c)} {\ln(1+c) - c (1+c)^{-1}} \\
  x & = & r / r_\Delta              \label{eq-x}  \; \; .
\end{eqnarray}
The two constants of integration, $g_1$ and $\rho_g$, are determined
by the two boundary conditions of the self-consistent solution.
First, the total gas mass within the dimensionless shock radius 
$x_{\rm ac} = r_{\rm ac} / r_\Delta$ must equal the total mass 
of accreted gas $f_b M_\Delta$.  Second, the ram pressure of 
accreting gas at $x_{\rm ac}$ must equal $P(x)$.  Furthermore,
the entropy just within the shock radius must be consistent
with the assumed polytropic relation.

In the case of cold accretion, entropy consistency and the 
pressure-balance boundary condition are satisfied if 
$T(x_{\rm ac}) = \mu m_p v_{\rm ac}^2 / 3$ and 
$\rho(x_{\rm ac}) = 4 \rho_1$.  The temperature
condition leads to
\begin{equation}
 g_1 = \left[ \frac {2} {3} (2 - x_{\rm ac}) -
       \frac {2 (\gamma_{\rm eff} - 1)} {\gamma_{\rm eff}} F(c_\Delta)
              \frac {\ln (1+c_\Delta x_{\rm ac})} {\ln (1+c)}
          \right] x_{\rm ac}^{-1} \; \; .
\end{equation}
For typical cluster parameters of $c_\Delta \approx 8$ and
$\gamma_{\rm eff} \approx 1.2$, we find $g_0(1) \approx 0.56$
and $g_1 \approx 0.11$ when $r_{\rm ac} = r_\Delta$.  
Furthermore, we find $g_1 = 0$ when $r_{\rm ac} \approx 1.11 r_\Delta$.
Because $g_0(x)$ declines with radius, we always have $g_0 \gg g_1$ 
for $x \lesssim x_{\rm ac}$; therefore, we elect to construct 
an approximate solution by setting $g_1 = 0$.

To apply the gas-mass boundary condition, we take advantage of the 
density condition at $x_{\rm ac}$. 
Because we have set $g_1 = 0$, we can then write the gas-mass 
boundary condition as
\begin{equation}
  \left( \frac {f_b \Delta \rho_{\rm cr}} {4 \rho_1} \right)
        =  3  \int_0^{x_{\rm ac}} 
                      \left[ \frac {g_0(r)} {g_0(r_{\rm ac})} 
                                   \right]^{1/(\gamma_{\rm eff}-1)}
                                   x^2 dx \\
\end{equation}
Defining $c_{\rm ac} = c_\Delta r_{\rm ac} / r_\Delta$ and recalling
that $\xi = 1 - x_{\rm ac}/2$, we therefore arrive at 
\begin{equation}
   \frac {(1-\xi_{\rm poly})^3} {\xi_{\rm poly}} = 
        [6 \tilde{\rho} I(\gamma_{\rm eff},c_{\rm ac})]^{-2} \; \; ,
\end{equation} 
for the polytropic model, with
\begin{equation}
  I(\gamma,c) \equiv  \int_0^1
                      \left[ \frac {\ln (1+cy)} {y \ln(1+c)} 
                      \right]^{1/(\gamma-1)} y^2 dy  \; \; .
\end{equation}
and
\begin{equation}
  \tilde{\rho} \equiv \frac {4} {3}
                      \left( \frac {2} {\Delta} \right)^{1/2}
                      (Ht)^{-1}
                      \frac {d \ln M} {d \ln t} \; \; ,
\end{equation}
so that $4 \rho_1 = (f_b \Delta \rho_{\rm cr}) 
(2 \xi)^{-1/2} x_{\rm ac}^{-3/2} \tilde{\rho}$.
Numerical integration shows that the expression $I(1.2,c) 
\approx (c/8) + 0.45$ is an excellent approximation in the 
interesting range $4 < c < 15$.
Note that the dependence of $\xi$ on the accretion rate is
encapsulated in $\tilde{\rho}$ and the dependence on internal
cluster structure is encapsulated in $I(\gamma,c)$.

The situation changes if the intracluster medium is
isentropic with an entropy level $K_1$ that is not determined
by the shock at the boundary.  For example, when preheated
gas with $K_1 > K_{\rm sm}$ accretes onto the cluster 
through a weak shock, the entropy internal to the cluster 
is determined primarily by the amount of preheating. 
In the isentropic case, we can set $\gamma_{\rm eff} = 5/3$, 
yielding an internal pressure
\begin{equation}
  P(x) = \left( \frac {T_\Delta f_b \Delta \rho_{\rm cr}} {\mu m_p}
                                               \right)
          \left( \frac {K_1} {K_\Delta} \right)^{-3/2}
          [g(x)]^{5/2} \; \; ,
\end{equation}
where $K_\Delta = T_\Delta (\mu m_p)^{-1} (f_b \Delta 
\rho_{\rm cr})^{-2/3}$.  At the bounding radius $x_{\rm ac}$
this pressure must equal the accretion pressure, implying that
\begin{equation}
 g(x_{\rm ac}) =  \left( \frac {2 \tilde{\rho}} {3} \right)^{2/5}
                  \left( \frac {K_1} {K_\Delta} \right)^{3/5}
                      (2-x_{\rm ac})^{1/5} x_{\rm ac}^{-1} \; \; .
\end{equation}
We can therefore write
\begin{equation}
 g(x) = \frac {4 F(c_\Delta)} {5}  
              \left[ \frac {\ln (1+c_\Delta x)} {x \ln(1+c_\Delta)} 
                 - \frac {\ln (1+c_\Delta x_{\rm ac})} 
                         {x_{\rm ac} \ln (1+c_\Delta)} 
                 + Q \right]    \; \; ,
\end{equation}
where
\begin{equation}
 Q \equiv \frac {5} {4 F(c_\Delta)}  
                     g(x_{\rm ac})  \; \; .
\end{equation}

In order to apply the gas-mass boundary condition, we need to
integrate $g^{3/2} x^2 dx$.  We can approximate this integral by
noting that $\ln (1+cx) /x \ln (1+c) \approx x^{-2/3}$ at 
$x \gtrsim 0.3$ to within 10\% for $c \sim 10$.  Because the
outer regions of an isentropic cluster dominate the overall
mass, the gas-mass boundary condition then becomes
\begin{equation}
  \left( \frac {K_1} {K_\Delta} \right)^{3/2} \approx
    3 \left[ \frac {4 F(c_\Delta)} {5} \right]^{3/2}
    x_{\rm ac}^2 J(Q) \; \; ,
\end{equation}
where
\begin{equation}
 J(Q) \equiv \int_0^1 (y^{-2/3} -1 +Q)^{3/2} y^2 dy \; \; .
\end{equation}
In the limit $Q \ll 1$, corresponding to a zero-pressure boundary,
the integral can be done analytically to give $J(Q) = [3 \Gamma(3)
\Gamma(5/2) ] / [2 \Gamma(11/2)] = 0.0762$.  In the limit $Q \gg 1$, 
corresponding to a constant-pressure interior, the integral
simplifies to $\frac {1} {3} Q^{3/2}$.  However, the most relevant
range for our purposes is $Q \sim 1$, because we are interested in 
clusters whose shock-generated entropy is comparable to that
produced by preheating. 
For these intermediate values of $Q$, numerical integration
shows that $J(Q) \approx Q/2$ is a decent approximation.  Applying
this approximation produces
\begin{equation}
  (1-\xi_{\rm isen}) \xi_{\rm isen}^{3/25} \approx
    2^{-28/25} \left( \frac {2} {3} \right)^{3/5}
    \left[ \frac {5} {4 F(c_\Delta)} \right]^{3/10}
    \left( \frac {2 \tilde{\rho}} {3} \right)^{-6/25}
                  \left( \frac {K_1} {K_\Delta} \right)^{27/50}
\end{equation}

As an accreting cluster makes the transition from the isentropic to 
the shock-dominated regime, we need a hybrid solution.
Let $f_K(t)$ be the fraction of accreted gas that is isentropic
at time $t$, let $x_K$ be the dimensionless radius within
which the gas is isentropic, and define $g_K = g(x_K)$.
The gas-mass condition for the polytropic 
($\gamma_{\rm eff} = 1.2$) region at $x > x_K$ 
is then
\begin{equation}
  (1-f_K) \left( \frac {f_b \Delta \rho_{\rm cr}} {4 \rho_1} \right)
        =  3  \int_{x_K}^{x_{\rm ac}} 
                      \left( \frac {g} {g_{\rm ac}} 
                                   \right)^5
                                   x^2 dx  \; \; ,
\end{equation}
where $g_{\rm ac} = g(x_{\rm ac})$.
The corresponding gas-mass condition for the isentropic region
is
\begin{equation}
  f_K \left( \frac {f_b \Delta \rho_{\rm cr}} {4 \rho_1} \right)
        =  3  \left( \frac {g_K} {g_{\rm ac}} \right)^5
              \int_0^{x_K} \left( \frac {g} {g_K} \right)^{3/2} x^2 dx  \; \; .
\end{equation}
Combining these expressions yields an expression for $\xi$ analogous
to that for the pure polytropic case:
\begin{equation}
   \frac {(1-\xi)^3} {\xi} = 
      \frac {1} {(6 \tilde{\rho})^2}
        \left[ \int_{y_K}^1 \left( \frac {g} {g_{\rm ac}} \right)^5 y^2 dy
         + \left( \frac {g_K} {g_{\rm ac}} \right)^5
          \int_0^{y_K} \left( \frac {g} {g_K} \right)^{3/2} y^2 dy
          \right]^{-2} \; \; ,
\end{equation} 
with $y = x / x_{\rm ac}$.  The first integral approaches
$I(1.2,c)$ as $x_K \rightarrow 0$, as long as $g \approx g_0$.  
To estimate the second integral, we note that $g_K \approx g_0(x_K)$ 
at $x \geq x_K$ implies that 
\begin{equation}
 g(x) \approx \frac {12} {5} g_K 
            \left[ \left( \frac {x} {x_K} \right)^{-2/3} 
                    - 1 + \frac {5} {12} \right] \; \; ,
\end{equation}
for $x \leq x_K$.  Thus, the second integral approaches 
$\approx (12/5)^{3/2} J(5/12)$ as $x_K \rightarrow x_{\rm ac}$.
We therefore use the following approximation to compute $\xi$
for the transitional case:
\begin{equation}
   \frac {(1-\xi_{\rm trans})^3} {\xi_{\rm trans}} \approx 
      \frac {1} {(6 \tilde{\rho})}^2    
      \left[ (1-f_K) \cdot I(1.2,c) 
            + f_K \cdot (12/5)^{3/2} J(5/12) \right]^{-2} \; \; ,
\end{equation} 

The solid lines in Figure~\ref{xi_tn} show $\xi = \min (\xi_{\rm poly},
\xi_{\rm trans}, \xi_{\rm isen})$ determined using the same 
cluster parameters that Tozzi \& Norman (2001) used to compute 
entropy models for clusters of $10^{14} \, h^{-1} \, M_\odot$ 
and $10^{15} \, h^{-1} \, M_\odot$.  We converted their $dM/dz$ 
relations for $\Lambda$CDM to $dM/dt$ relations, yielding 
$M \propto t^{0.931 - 1.0 \log H_0 t}$ for the
$10^{15} \, h^{-1} \, M_\odot$ cluster and
$M \propto t^{0.547 - 0.914 \log H_0 t}$ for the
$10^{14} \, h^{-1} \, M_\odot$ cluster.
In order to reproduce the time-dependence of the concentration
parameter we interpolated a power law in $(1+z)$ between 
their mass-concentration relations at $z = 0$ and $z = 1$, 
giving $c_\Delta = 8.5 (M_\Delta / 10^{15} \, h^{-1} \, M_\odot)^{-0.086} 
(1+z)^{-0.65}$.  We also neglected the difference between 
$c_{\rm ac}$ and $c_\Delta$ in computing $I(\gamma,c)$.  
Note that this model is not a good representation when
$\xi \ll 0.5$ because it assumes that the time when gas reaches
the shock front is twice the time it took to reach its turnaround
radius.

%\begin{appendix}
\begin{center}
APPENDIX B\\ ~ \\
INHOMOGENEOUS ACCRETION AND ENTROPY PRODUCTION \\
\end{center}

\renewcommand{\theequation}{B\arabic{equation}}
\setcounter{equation}{0}

The following computation
explicitly demonstrates that inhomogenous accretion 
produces less entropy than homogeneous accretion. Define a
dimensionless baryon density $x \equiv 4 \pi r^2 v \rho /
\dot{M}$ and let $f(x) dx$ be the fraction of the 
accreting volume with density between $x$ and $x + dx$,
so that $\int x f(x) dx = 1$.
The fraction of accreting {\em mass} with density in
this range is then $x f(x) dx$, because of how density
is defined.  Thus, the mean mass-weighted density 
exceeds the mean volume-weighted density by a factor
\begin{equation}
 \int x^2 f(x) dx = \int (x-1)^2 f(x) dx + 1 \; \; .
\end{equation}
In the homogeneous case, $f(x)$ is a delta function
at $x=1$, and the mass-weighted density equals the
volume-weighted density.  However, if there is any
inhomogeneity, the integral on the right-hand side will 
be greater than zero, implying that the mass-weighted
mean density must be larger than the volume-weighted 
mean density.

We are interested in the mean mass-weighted entropy of
gas accreting at time $t$.  If all the accreting gas
moves with the same velocity and passes through an accretion 
shock at the same radius, then the mean mass-weighted
entropy of accreting gas is
\begin{equation}
\label{eq-meank}
 \bar{K}(t) = K_{\rm sm}(t) \int x^{-2/3} \cdot x f(x) dx
            = K_{\rm sm}(t) \int x^{1/3} f(x) dx \; \; .
\end{equation}
We define $y = x^{1/3}$ and $g(y) dy = f(x) dx$, so that 
$\int y^3 g(y) dy$ = $\int g(y) dy = 1$, implying 
$\int (y^3 -1) g(y) dy = 0$.
This equation leads to
\begin{equation}
 \int (y-1) g(y) dy = - \frac {1} {3} \int (y+2)(y-1)^2 g(y) dy \; \; .
\end{equation}
Because the integrand on the right-hand side is always
positive, the integral on the left-hand side must be
less than zero, except in the homogeneous case.  Thus,
we have
\begin{equation}
  \frac {\bar{K}(t)} {K_{\rm sm}(t)} = \int x^{1/3} f(x) dx
                   = \int y \, g(y) dy \leq 1 \; \; ,
\end{equation}
demonstrating that homogeneous accretion maximizes entropy
production in an accretion shock.  Notice that this conclusion
also applies to smoothing of the flow in solid angle or in
time, meaning that the flow must also be isotropic and steady
in order to maximize the postshock entropy.

\vspace{2em}
%\clearpage

%\end{appendix}


\clearpage


\begin{thebibliography}{}

\bibitem{ae99}
Arnaud, M., \& Evrard, A. E. 1999, \mnras, 305, 631

\bibitem{bblp02}
Babul, A., Balogh, M. L., Lewis, G. F., \& Poole, G. B. 2002, \mnras, 330, 329

\bibitem{bbp99}
Balogh, M. L., Babul, A., \& Patton, D. R. 1999, \mnras, 307, 463

\bibitem{b85}
Bertschinger, E. 1985, \apjs, 58, 39

\bibitem{bem01}
Bialek, J. J., Evrard, A. E., \& Mohr, J. J. 2001, \apj, 555, 597

\bibitem{bcek91}
Bond, J. R., Cole, S., Efstathiou, G., \& Kaiser, N. 1991, \apj, 379, 440

\bibitem{bgwmtlqs01}
Borgani, S., Governato, F., Wadsley, J., Menci, N., Tozzi, P., Lake,
G., Quinn, T., \& Stadel, J. 2001, \apjl, 559, L71

\bibitem{bgwmtqsl02}
Borgani, S., Governato, F., Wadsley, J., Menci, N., Tozzi, P., 
Quinn, T., Stadel, J., \& Lake, G., 2002, \mnras, 336, 409

\bibitem{b91}
Bower, R. G. 1991, \mnras, 248, 332

\bibitem{b97}
Bower, R. G. 1997, \mnras, 288, 355

\bibitem{bblbcf01}
Bower, R. G., Benson, A. J., Lacey, C. G., Baugh, C. M., Cole, S., \&
Frenk, C. S. 2001, \mnras, 325, 497

\bibitem{b99}
Bryan, G. L. 1999, Computing in Science \& Engineering, 1999, 1, 46

\bibitem{b00}
Bryan, G. L. 2000, \apj, 544, L1

\bibitem{bv01}
Bryan, G. L., \& Voit, G. M. 2001, \apj, 556, 590

\bibitem{bksskkpd01}
Bullock, J. S., Kolatt, T. S., Sigad, Y., Somerville, R. S.,
Kravtsov, A. V., Klypin, A. A., PRimack, J. R., \& Dekel, A.
2001, \mnras, 321,559

\bibitem{cff78} 
Cavaliere, A., \& Fusco-Femiano, R. 1976, \astap, 49, 137
 
\bibitem{cmt97}
Cavaliere, A., Menci, N., \& Tozzi, P. 1997, \apj, 484, L21

\bibitem{cmt98}
Cavaliere, A., Menci, N., \& Tozzi, P. 1998, \apj, 501, 493
 
\bibitem{cmt99}
Cavaliere, A., Menci, N., \& Tozzi, P. 1999, \mnras, 308, 599

\bibitem{dm02}
De Grandi, S., \& Molendi, S. 2002, \apj, 567, 163

\bibitem{dd02} 
Dos Santos, S., \& Dor\'e, O. 2002, \astap, 383, 450

\bibitem{es91}
Edge, A. C., \& Stewart, G. C. 1991, \mnras, 252, 414

\bibitem{enf98}
Eke, V., Navarro, J. F., \& Frenk, C. S. 1998, \apj, 503, 569

\bibitem{ens01}
Eke, V., Navarro, J. F., \& Steinmetz, M. 2001, \apj, 554, 114

\bibitem{ef99}
Ettori, S., \& Fabian, A. C. 1999, \mnras, 305, 834

\bibitem{eh91}
Evrard, A. E., \& Henry, J. P. 1991, \apj, 383, 95

\bibitem{f99_sbclust}
Frenk \etal 1999, \apj, 525, 554

\bibitem{frb01}
Finoguenov, A., Reiprich, T. H., \& B\"ohringer, H. 2001, \astap, 368, 749

\bibitem{fjbp02}
Finoguenov, A., Jones, C., B\"ohringer, H., Ponman, T. J. 2002,
\apj, 578, 74

\bibitem{glrb02}
Gomez, P., Loken, C., Roettiger, K., \& BUrns, J. O. 2002,
\apj, 569, 122

\bibitem{hp00} 
Helsdon, S. F., \& Ponman, T. J. 2000, \mnras, 315, 356

\bibitem{k96}
Kaiser, N. 1986, \mnras, 222, 323

\bibitem{k91}
Kaiser, N. 1991, \apj, 383, 104

\bibitem{kp97}
Knight, P. A., \& Ponman, T. J. 1997, \mnras, 289, 955

\bibitem{ks01}
Komatsu, E., \& Seljak, U. 2001, \mnras, 327, 1353

\bibitem{lc93}
Lacey, C., \& Cole, S. 1993, \mnras, 262, 627

\bibitem{ll}
Landau, L. D., \& Lifshitz, E. M. 1959, Fluid Mechanics (London: Pergamon)

\bibitem{m98} 
Markevitch, M. 1998, \apj, 504, 27

\bibitem{m00}
Markevitch, M. \etal 2000, \apj, 541, 542

\bibitem{mfsv98} 
Markevitch, M., Forman, W. R., Sarazin, C. L., \& Vikhlinin, A. 1998, 
\apj, 503, 77

\bibitem{mvfs99} 
Markevitch, M., Vikhlinin, A., Forman, W. R., \& Sarazin, C. L. 1999, 
\apj, 503, 77

\bibitem{mfls03}
Mushotzky, R., Figueroa-Feliciano, E., Loewenstein, M., \& Snowden, S. L.
2003, astro-ph/0302267

\bibitem{nfw95}
Navarro, J. F., Frenk, C. S., \& White, S. D. M. 1995, \mnras, 275, 720

\bibitem{nfw97}
Navarro, J. F., Frenk, C. S., \& White, S. D. M. 1997, \apj, 490, 493

\bibitem{nw93}
Navarro, J. F., \& White, S. D. M. 1993, \mnras, 265, 271

\bibitem{nb98} 
Norman, M. L., \& Bryan, G. L. 1998, Numerical Astrophysics 1998, 
ed. S. Miyama \& K. Tomisaka, (Dordrecht: Kluwer), p. 19

\bibitem{pcn99} 
Ponman, T. J., Cannon, D. B., \& Navarro, J. F. 1999, Nature, 397, 135

\bibitem{pa03}
Pratt, G., \& Arnaud, M. 2003, astro-ph/0304017

\bibitem{sd93}
Sutherland, R. S., \& Dopita, M. A. 1993, \apjs, 88, 253

\bibitem{tn01}
Tozzi, P., \& Norman, C. 2001, \apj, 546, 63

\bibitem{vmm01} 
Vikhlinin, A., Markevitch, M., Murray, S. S. 2001, \apj, 551, 160

\bibitem{vb01}
Voit, G. M., \& Bryan, G. L. 2001, Nature, 414, 425

\bibitem{vbbb02}
Voit, G. M., Bryan, G. L., Balogh, M. L., \& Bower, R. G. 2002, 
\apj, 576, 601

\bibitem{vd98}
Voit, G. M., \& Donahue, M. 1998, \apj, 500, L111

\bibitem{wfn00}
Wu, K. K. S., Fabian, A. C., \& Nulsen, P. E. J. 2000, \mnras, 318, 889

\bibitem{wx02a}
Wu, X.-P., \& Xue, Y.-J. 2002a, \apj, 569, 112

\bibitem{wx02b}
Wu, X.-P., \& Xue, Y.-J. 2002b, \apj, 572, 19


\end{thebibliography}
\end{document}